\documentstyle[11pt,aaspp4]{article}
\lefthead{Rix et al.}
\righthead{Dynamical modeling of velocity profiles}

\def\Reff{R_{\rm eff}}
\def\rc{r_{\rm c}}
\def\Np{N_{\rm p}}
\def\Nk{N_{\rm k}}
\def\Nc{N_{\rm c}}
\def\No{N_{\rm o}}
\def\vt{v_{\rm t}}
\def\Vc{V_{\rm c}}
\def\rh{r_{\rm h}}

\def\etal{{\it et al.~}}
\def\eg{e.g.,~}

\def\asec{^{\prime\prime}}

\def\kms{{\rm\,km\,s^{-1}}}

\def\Mpc{{\rm\,Mpc}}

\def\fr#1#2{{#1\over#2}}

\def\piby2{{\pi \over 2}}

\def\d{{\rm d}}
\def\VP{{\rm VP}}

\def\spose#1{\hbox to 0pt{#1\hss}}
\def\lta{\mathrel{\spose{\lower 3pt\hbox{$\sim$}}
    \raise 2.0pt\hbox{$<$}}}
\def\gta{\mathrel{\spose{\lower 3pt\hbox{$\sim$}}
    \raise 2.0pt\hbox{$>$}}}

\begin{document}

\title{Dynamical Modeling of Velocity Profiles:\\
       The Dark Halo Around the Elliptical Galaxy NGC~2434}

\author{Hans-Walter Rix,\altaffilmark{1,2,3}
        P.~Tim de Zeeuw,\altaffilmark{4}
        C.~Marcella Carollo,\altaffilmark{5,4,7}\\
        Nicolas Cretton\altaffilmark{4} \& 
        Roeland P.~van der Marel\altaffilmark{6,7}}

\altaffiltext{1}{Steward Observatory, University of Arizona, Tucson, 
                 AZ 85721}
\altaffiltext{2}{Max-Planck-Institut f\"ur Astrophysik,
                 Karl-Schwarzschild-Strasse 1, Garching 87540, Germany}
\altaffiltext{3}{Alfred P.~Sloan Fellow}
\altaffiltext{4}{Sterrewacht Leiden, Postbus 9513, 2300 RA Leiden, 
                 The Netherlands}
\altaffiltext{5}{Department of Physics and Astronomy, 
                 Johns Hopkins University, Baltimore, MD 21218}
\altaffiltext{6}{Institute for Advanced Study, Olden Lane, Princeton, 
                 NJ 08540}
\altaffiltext{7}{Hubble Fellow}

%%%%%%%%%%%%%%%
% Abstract
%%%%%%%%%%%%%%%

\clearpage
\begin{abstract}
We describe a powerful technique to model and interpret the stellar
line-of-sight velocity profiles of galaxies. It is based on
Schwarzschild's approach to build fully general dynamical models.  A
representative library of orbits is calculated in a given potential,
and the non-negative superposition of these orbits is determined that
best fits a given set of observational constraints.  Our implementation
incorporates several new features:
(i) we calculate velocity profiles and
represent them by a Gauss-Hermite series. This allows us
to constrain the orbital anisotropy in the fit. 
(ii) we take into account the error on each observational constraint
to obtain an objective $\chi^2$ measure for the quality-of-fit.
Given
the observational constraints, the technique assesses the relative
likelihood of different orbit combinations in a given potential, and
of models with different potentials. In our implementation only
projected, observable quantities are included in the fit,
aperture binning and seeing convolution 
of the data are properly taken into
account, and smoothness of the models in phase-space can be enforced
through regularization. This scheme is valid for any 
geometry.\looseness=-2

%Modeling the velocity profile shapes is very important; they constrain
%the anisotropy of the velocity distribution, which is the main
%uncertainty in estimating the gravitational potential from
%observational data. To illustrate the use of velocity profile shapes,
%we model pseudo-data generated from an isotropic Hernquist model. We
%we show that the potential is much more tightly constrained by the
%data if not only the velocity dispersion is included in the fit, but
%also the Gauss-Hermite moment $h_4$.

In a first application of this method,
we focus here on spherical geometry; axisymmetric modeling is described
in companion papers by Cretton \etal and van~der~Marel \etal 
We test the scheme on pseudo-data drawn from an isotropic Hernquist model, 
and then apply it to the issue of dark halos around elliptical galaxies.
We model radially extended
stellar kinematical data for the E0 galaxy NGC~2434, obtained
by Carollo et al.
~Constant mass-to-light ratio models are clearly ruled out,
regardless of the orbital anisotropy. To study the amount of dark
matter needed to match the data, we considered a sequence of
cosmologically motivated `star+halo' potentials. These potentials
are based on the CDM simulations by Navarro et al., but also account
for the accumulation of baryonic matter; they are specified by
the stellar mass-to-light ratio $\Upsilon_{*,B}$ and the
characteristic halo velocity, $V_{200}$.
The star+halo models provide an excellent fit to the
data, with $\Upsilon_{*,B}=3.35\pm0.25$ (in B-band solar units)
and $V_{200}=450\pm100$~km/s.
The best-fitting potential has a circular velocity $\Vc$ that is
constant to within $\sim 10\%$ between $0.2$--3 effective radii and
is very similar to the best-fitting
logarithmic potential, which has $\Vc = 300 \pm 15 \kms$. In NGC 2434
roughly half of the mass within an effective radius is dark. Models
without a dark halo overestimate the mass-to-light ratio of the
stellar population by a factor of $\sim 2$.
\end{abstract}

%%%%%%%%%%%%%%%
% Keywords
%%%%%%%%%%%%%%%

% The different journals have different requirements for keywords.  The
% keywords.apj file, found on aas.org in the pubs/aastex-misc directory, 
% contains a list of keywords used with the ApJ and Letters.  These are 
% usually assigned by the editor, but authors may include them in their 
% manuscripts if they wish. 


%%%%%%%%%%%%%%%
% Beginning of main text
%%%%%%%%%%%%%%%

\clearpage
\section{Introduction}

Mapping how the ratio of luminous to dark matter in a galaxy changes
as a function of radius provides an important test for galaxy
formation scenarios.  Numerical simulations of halo formation in a
cosmological context have reached a level where they can predict the
radial profile of an isolated dark halo (Navarro \etal 1996,
hereafter NFW; Cole \& Lacey 1996).
This profile is altered by the presence of
dissipative, baryonic matter, which collects at the center and
contracts the dark matter profile. This contraction may provide a
natural explanation for the observed fact that the circular velocity
is approximately constant with radius in spiral galaxies (Blumenthal
\etal 1986; NFW). Elliptical galaxies are more centrally concentrated than
spiral galaxies of the same mass, suggesting that they may have
circular velocities that are higher in the inner parts than in the
outer parts.

Observational studies of the dark halos in spiral galaxies (e.g.,
van Albada \etal 1985) are comparably straightforward.  HI gas
provides an excellent tracer to large radii. To interpret the
kinematics it is justifiable to assume a nearly co-planar distribution
with nearly circular orbits, upon which the gravitational potential
can be constrained from the observables. By contrast, the majority 
of elliptical galaxies do not have HI disks that are in
equilibrium, and the transition from where the
stellar mass dominates to where the dark halo dominates has remained
poorly constrained. This has been due to the difficulties in obtaining
unambiguous results from the stellar kinematics, as caused by two main
problems. First, modeling the stellar dynamics for ellipticals is much
more complex than for the cold disks of spirals (\eg de~Zeeuw \& Franx
1991; Bertin \& Stiavelli 1993; de Zeeuw 1996). Random motions
dominate, and the stars can occupy a host of qualitatively different
orbits in any given potential. Hence, the dynamical modeling must
solve for {\it both} the potential {\it and} the orbital distribution
of the stars, given the observed projected positions and velocities of
stars. In practice, most existing studies have not done this. Instead,
the orbital structure has often been assumed a priori, by requiring
that the distribution function (hereafter, DF; i.e., the number of
stars per unit volume of the phase space of stellar positions and
velocities) has a certain simple form. The second reason for the poor
understanding of the star--halo connection in ellipticals has been the
fact that until recently good stellar kinematic data were available
only out to approximately one effective radius $\Reff$. This left more
than half the stellar mass kinematically unconstrained. In addition,
the data were generally restricted to measurements of the two lowest
order velocity moments, i.e., the mean streaming velocity $V$ and the
line-of-sight velocity dispersion $\sigma$. These quantities contain
no or little independent information on the intrinsic velocity
dispersion anisotropy of the system (Binney \& Mamon 1982), which
provides the main indeterminacy in the modeling.
As a consequence, stellar dynamical indications for the 
presence of dark halos around elliptical galaxies have remained
ambiguous. There have been indications for dark halos in studies that
employed restricted classes of dynamical models (e.g., van der Marel
1991; Saglia \etal 1992). However, the few models in the literature
that had the full freedom of rearranging orbits (e.g., Richstone \&
Tremaine 1984; Dejonghe 1989) have been able to fit the data for most
ellipticals without requiring any dark matter (Katz \& Richstone 1985;
Saglia \etal 1993; Bertin \etal 1994). Discrete kinematical tracers in
elliptical galaxies (i.e., planetary nebulae and globular clusters)
have the advantage that they can be observed to larger radii, but they
have the disadvantage of small-number statistics. 
Dynamical modeling of their observed radial velocities is beset
by the same degeneracies that plague the interpretation of the
integrated light measurements. Not surprisingly, these studies
have produced similarly ambiguous results
(\eg Ciardullo, Jacoby \& Dejonghe 1993; Tremblay,
Merritt \& Williams 1995). 
Yet, as summarized in various reviews
(e.g., Ashman 1992; de Zeeuw 1995; Saglia 1996), there is
independent evidence for extended dark matter from X-ray
measurements of luminous galaxies (\eg Forman, Jones \& Tucker 1985;
Awaki et al.\ 1994), from HI kinematics (Franx \etal 1994),
and from gravitational lensing (\eg Maoz \& Rix 1993;
Kochanek 1995). It is likely that this apparent discrepancy is
attributable to the shortcomings of the stellar dynamical tests,
i.e., to the uncertainty about the orbital distributions resulting
from insufficiently constraining data.

There are several reasons for being optimistic that this situation can
be improved and that the luminous to dark matter distribution in
elliptical galaxies can now be investigated in some detail through
stellar dynamical studies. First, improved CCD technology (larger
detector size, higher QE, lower read-out noise and dark current),
combined with improved strategies to ensure accurate sky subtraction
and better stellar template matching, have made it possible to obtain
stellar kinematics to 2--4 $\Reff$ (Carollo \etal 1995, hereafter C95;
Statler \etal 1996). With data reaching out to these radii, kinematic
constraints can be obtained over most of the stellar body. Second,
improved analysis techniques (e.g., Rix \& White 1992; van~der~Marel
\& Franx 1993, hereafter vdMF; Kuijken \& Merrifield 1993)
make it possible now to extract the
entire line-of-sight velocity profile (VP) from the absorption-line
spectra, rather than only its lowest order moments $V$ and $\sigma$.
The higher moments of the VP, which can now be measured, contain
essential information about the anisotropy of the velocity
distribution (\eg Dejonghe, 1987; Gerhard 1993; C95).

The main purpose of this paper is to describe the development
and application 
of theoretical tools that permit to model
and fit VP shape data in a flexible and objective way. 
In its most general form the modeling technique presented here can answer the
question: given a variety of gravitational potentials and given a set
of observational constraints (photometry and kinematics, including
VPs), what is the relative likelihood of the different potentials? For
each potential the orbital distribution is determined that best fits
the data, and the likelihood follows from the quality of the fit. Thus
the method allows one to determine which are the best-fitting
potentials, and which potentials are excluded by the data.

Our technique is based on the numerical calculation of a
representative library of orbits in a chosen potential, and the
subsequent determination of the non-negative superposition of these
orbits that best fits the data. This approach was pioneered by
Schwarzschild (1979), who required the orbit superposition to
reproduce the galaxy density, and so built triaxial galaxy models.
Richstone (1980; 1984) built scale--free
axisymmetric models with this technique. In the past decade, the
approach has been used to build a variety of spherical, axisymmetric
and triaxial galaxy models, which also include the observed radial
velocities and/or velocity dispersions as constraints
(e.g., Pfenniger 1984; Richstone \& Tremaine 1984, 1985, 1988;
Levison \& Richstone 1985; Zhao 1996).
%Richstone \& Tremaine (1985) extended the approach by including
%observed velocity dispersion measurements as a constraint, under the
%assumption of a spherical geometry. Their method was used subsequently
%in many papers to study the presence of black holes in galactic nuclei
%(as reviewed by Kormendy \& Richstone 1995). Levison \& Richstone
%(1985) extended the approach to the scale--free axisymmetric case, and
%Zhao (1996) recently used a similar approach to construct tumbling
%triaxial models for the Galactic bar.

We have built on this previous work by adding two new 
features: (i) we calculate and compare an arbitrary number of moments
of the Gauss-Hermite series expansion of the VP, and show how they can
be used as linear constraints on the model; and (ii) we take into
account the error on each observational constraint in the
superposition procedure and hence obtain an objective measure for the
quality-of-fit. Only projected, observable quantities are included in
the fit. We include a proper seeing convolution of each orbit, so that
the observational setup of the data (including aperture binning)
can be accurately taken into account.
The addition of VP modeling not only
allows us to interpret the VP shape data that is now becoming
available, but it also removes the need for additional simplifications
and assumptions. We do not have to assume that the true lowest order
velocity moments of the models can be compared without bias to the
best Gaussian fits $V$ and $\sigma$ obtained from the data. This was
done in previous implementations, and is also implicit in modeling
based on the Jeans equations (e.g., Merritt \& Oh 1997). Avoiding this
assumption removes the possibility of systematic errors (easily
10--20\%; vdMF) in the interpretation of $V$ and $\sigma$. We also do
not assume that the local velocity distributions are Gaussian
everywhere along the line-of-sight.

Our extension of Schwarzschild's method is valid for any geometry.
To illustrate and test the new elements in the most straightforward
way, we restrict ourselves here to the spherical case, which
simplifies the calculation of the stellar orbits. The treatment of
the VP shapes and of the orbital superposition is fully general.
We test the method on analytic models, and then apply it to the
kinematic measurements for the E0 galaxy NGC 2434 obtained by C95;
preliminary results from this analysis are published in Rix (1996a,b).
The extension of our scheme to axisymmetric systems is described in
Cretton \etal (1997) and applied in van~der~Marel \etal (1997a,b).

%To illustrate and test the new elements of our technique in the most
%straightforward way, we restrict ourselves here to the spherical case.
%Other techniques to constrain the potentials of spherical stellar
%systems through modeling of observed VP shapes have been proposed
%e.g., by  Dejonghe \& Merritt (1992), Merritt (1993a),
%Merritt \& Saha (1993). Next to its
%flexibility and unparametrized nature, the primary strength of our
%technique is that it can be generalized to more complicated geometries
%in a straightforward manner. 
%
%Preliminary results from this spherical modeling are presented in Rix
%(1996ab); the extension to axisymmetric systems is
%described in Cretton \etal (1997) and applied in van~der~Marel \etal
%(1997a,b).

The paper is organized as follows. In Section~2 we describe the
modeling procedure. In Section~3 we test the method on isotropic
Hernquist models, and in Section~4 we use it to study the presence and
properties of a dark halo around the E0 galaxy NGC~2434. In the
analysis we use a family of cosmologically motivated galaxy
potentials that is discussed in Appendix~B. We summarize our
results in Section~5.
%
%The paper is organized as follows. In Section~2 we describe the
%modeling procedure. In Section~3 we use it to model the data
%(presented by C95) for the E0 galaxy NGC~2434, to study the presence
%and properties of a dark halo. In Section~4 we summarize the
%results. In the analysis we use a family of cosmologically motivated
%galaxy potentials that is discussed in Appendix~B. In Appendix~A we
%show how the DF of a model can be recovered from the orbital weights
%returned by the technique.

\section{Modeling technique}

Our extension of Schwarzschild's method tests in three steps whether a
given potential is compatible with all observational data:
\begin{enumerate}{}{\topsep=0pt\parsep=0pt\itemsep=0pt\listparindent=0pt}
\item A representative library of orbits is calculated in the 
      chosen potential.
\item Each orbit is projected onto the space of observables.
\item The combination of orbits with non-negative occupation numbers is 
      found that best matches the data, taking into account 
      the observational errors.
\end{enumerate}

For computational convenience, we here carry out steps 1 and 2 for
spherical geometry. Our description of step 3 is fully general.
Other techniques to constrain the potentials of spherical stellar
systems through modeling of observed VP shapes are available
(e.g., Dejonghe \& Merritt 1992; Merritt 1993a; Merritt \& Saha
1993), but these do require the availability of analytic integrals
of motion. Our approach has the advantage that it can be generalized
to more complicated geometries in a straightforward manner.

\subsection{Orbit Library}

In a spherical gravitational potential $\Phi(r)$, all orbits are
planar rosettes, characterized by four isolating integrals of motion:
the energy $E$, and the three components of the angular momentum $\vec
L$. For each energy, $L=|\vec{L}|$ lies in the interval $[0, L_{\rm
max}]$, where $L_{\rm max}$ is the angular momentum for the circular
orbit at energy $E$. This circular orbit has a radius $\rc$, given by
the implicit equation
\begin{equation}
   \Phi (\rc) + {\textstyle {1 \over 2}} \rc
   \left [ {{\partial\Phi}\over{\partial r}} \right ]_{r=\rc} = E.
\end{equation}
To cover phase space, we choose a grid in the $(E,L)$ plane (the
Lindblad diagram) in the following way. We specify a set of radii
$r_{{\rm c},i}$, $i=1, \ldots, N_E$, which are spaced logarithmically
at small radii (corresponding to observed radii $<0.5\asec$) and
linearly at large radii with $\Delta \rc\sim 1\asec$ (the typical
resolution of the observational constraints). Each radius $r_{{\rm
c},i}$ defines an energy grid point $E_i$, with an associated $L_{{\rm
max},i}$. For each of the $N_E$ grid points $E_i$, we choose $N_L$
angular momenta $L_{ij} \equiv \beta_j L_{{\rm max},i}$, $j=1, \ldots,
N_L$. The numbers $\beta_j$ are distributed uniformly in the interval
$[\epsilon, 1-\epsilon]$, with $\epsilon=0.02$. This choice of
$\epsilon$ excludes the exactly radial and circular orbits, which
avoids numerical difficulties in the radial orbit calculation for
singular potentials, and avoids the sharp edges in the projected
density and VP of a circular orbit.  Typically we set $N_E \approx 50$
and $N_L \approx 10$.

The orbits in a spherical model need only be integrated in the radial
dimension, which provides the instantaneous radius $r$ and radial
velocity $v_r$ (\eg Binney \& Tremaine 1987). The tangential velocity
$\vt$ follows from $\vt = L/r$.  Each orbit is started at its
apocenter, and is calculated for an integral number of periods. For a
spherical potential one half of a period is in principle sufficient,
but calculating a small number of complete periods ($\sim 5$) reduces
discreteness effects in the result, arising from the finite number of
time-steps. We chose approximately 300 time steps per radial period.
During the integration the time steps were adjusted to conserve
energy equally well at each step. For every orbit the
fractional energy conservation was at least $10^{-5}$ over the whole
integration.

Our orbit integrations are carried out with a simple
predictor-corrector integrator, because the orbits need not be
computed with great accuracy. If, for example, the orbit were to drift
in energy by a small amount during the integration, the orbit would no
longer represent a $\delta$-function in phase space, but a short line
or small area in the $(E,L)$ plane. But since the (presumably)
continuous DF is represented by a finite number of discrete orbits (or
`basis vectors'), there is no reason to prefer true $\delta$-functions
over `fuzzy' ones.  Also, physical meaning cannot be attached to rapid
fluctuations of the phase space density, given realistically available
constraints. Indeed, we have found it advantageous to assemble each
orbit $(E_{i},L_{ij})$ from a number of `sub-orbits', typically about
25, whose integrals of motion are drawn at random from a small phase
space cube $(\delta E, \delta L)$ around $(E_i, L_{ij})$ (see also
Zhao 1996). This reduces sharp edges in the projected density and VP
(especially for nearly circular orbits). It also relieves memory
requirements by reducing the number of orbits that must be stored.

For testing and debugging the algorithms it proved useful to construct
smoother building blocks, $f(E)$--{\it components}, in addition to the
$(E,L)$--{\it orbits}. They can be viewed as a weighted combination of
orbits with different $L$, but the same $E$. They can be constructed
almost analytically for any spherical potential, as described in
Appendix~A.  These components can be used to build isotropic models
with $f=f(E)$.  Such models provide a useful test case, because their
projected properties can often be calculated analytically. For
axisymmetric models, $f(E,L_z)$--components (with $z$ along the
symmetry axis) prove to be similarly useful (Cretton \etal 1997).

\subsection{Observables}

\subsubsection{Projected orbits}

We adopt a Cartesian coordinate system $(x,y,z)$, with the $z$-axis
directed towards the observer. The associated cylindrical and polar
coordinate systems are denoted $(R,z,\phi)$ and $(r,\theta,\phi)$.
This makes $R$ the projected radius in the $(x,y)$ plane of the
sky. The tangential velocity $\vt$ satisfies $\vt^2 = v_{\theta}^2 +
v_{\phi}^2$. An angle $\xi$ defines $v_{\theta}$ and $v_{\phi}$ in
terms of $\vt$, through $v_{\theta} = \vt \cos \xi$ and $v_{\phi} =
\vt \sin \xi$.

As an orbit is integrated, it is projected onto the space of the
observables $(x, y, v_z)$. In the following, we will refer to the
line-of-sight velocity $v_z$ as $v$. To store the projected orbital
properties we adopt a grid (i.e., a storage cube) in the
$(x,y,v)$-space. The spatial coordinate spacing is matched to the
pixel grid of the photometric or spectroscopic observations (\eg $\sim
1\asec$ for ground-based data). The velocity coordinate should cover
in principle $[-v_{\rm escape},v_{\rm escape}]$. We found in practice
that 30-50 velocity bins covering the range $[-4\sigma_{\rm
max},4\sigma_{\rm max}]$ is sufficient, where $\sigma_{\rm max}$ is
the largest observed velocity dispersion.

The orbit integration yields only the phase-space coordinates $(r,
v_r, \vt)$ at each time step. However, for projection onto the
space of the observables, all six phase-space coordinates are
required. These are obtained by drawing a random viewing angle and a
random direction of the tangential velocity vector at each time step,
i.e., we draw $\cos \theta \in [-1,1]$, $\phi \in [0,2\pi]$ and $\xi
\in [0,2\pi]$ from uniform distributions. This yields the following
observables at the given time step:
\begin{equation}
x = r \sin \theta \cos \phi , \qquad
y = r \sin \theta \sin \phi , \qquad
v = v_r \cos \theta - \vt \cos \xi \sin \theta .
\end{equation}
This procedure properly takes into account the fact that the storage
cube should contain the average contribution of {\it all} trajectories
that correspond to the given $(E_i,L_{ij})$ (i.e., it ensures that the
models have DFs that depend only on the modulus of ${\vec L}$, and not
on its direction). The procedure may be repeated numerous times at each
time step, to obtain several different $(x,y,v)$ for the same $(r,
v_r, \vt)$. This corresponds to viewing the model from all
geometrically equivalent angles at a fixed time, and creates a
smoother projection.

We denote the different orbits in the library (each corresponding to a
fixed combination $(E_i,L_{ij})$) with the index $k$, with
$k=1,\ldots,\No$. The total number of orbits $\No = N_E \times
N_L$. The {\it occupation weight} of orbit $k$ on the storage cube
cell centered on $(x,y,v)$ is denoted as $w^k_{xyv}$. At the start of
the integration of each orbit, the weights are set to zero. As the
integration of the trajectory proceeds, and projected coordinates are
obtained as described above, a weight is added to the cube cell that
contains the projected coordinates. This weight is chosen equal to the
average of the sizes of the previous and the next time-step in the
integration. This assignment of weights is effectively a Monte-Carlo
integration of the orbit over the grid, because the chance of dropping
a weight in a grid cell is proportional to the time the orbit spends
in it. Once the orbit calculation (and projections) are finished, the
occupation weights for each orbit are normalized to unit mass,
\begin{equation}
  \sum_{xyv} w^k_{xyv} = 1 , \qquad \forall k.
\end{equation}
Sometimes projected quantities extend beyond the boundaries of the
storage cube. In this case no weights are added to the grid, but the
time spent outside the cube is used in the normalization.

Figure~1 illustrates the occupation weights for four orbits in a test
model, all with the same energy $E$ but with different angular momenta
$L/L_{\rm max}$. Because of the spherical symmetry of the system, the
weights can be conveniently displayed in the two-dimensional space of
projected radius $R$ and line-of-sight velocity $v$\footnote{The use
of a three-dimensional $(x,y,v)$ storage cube in our technique is
motivated by the fact that realistic observational setups generally do
not have circular symmetry on the sky. It also makes the generalization
to axisymmetric systems simple. In axisymmetric or triaxial
potentials the orbit integration proceeds differently, but after
projection, the fitting of the observational
constraints through orbit superposition is identical.}. The
differences in the appearance of the orbits in this space of
observables is evident.

\subsubsection{PSF convolution}

It is important for model predictions to incorporate the observational
setup and to account for the point-spread-function (PSF) of the data.  The
final model is a linear super-position of orbits and the PSF
convolution of an orbit is also a linear operation.  Therefore, the
two operations commute and the PSF convolution can be carried out
separately for each orbit.  The PSF does not correlate
velocities\footnote{The finite spectrograph resolution, i.e., the PSF
in the velocity direction, is being accounted for by the kinematic
data analysis technique (\eg Rix \& White 1992; vdMF).} and hence is
carried out separately for each velocity slice of the storage
cube. Each velocity slice is an `iso-velocity' image of the orbit on a
Cartesian grid. Hence, the PSF convolution is most efficiently carried
out by a Fast Fourier Transform of these iso-velocity images.  As many
orbits, \eg the tightly bound ones, only occupy a small fraction of
the spatial grid, the size $2^N$ of the FFT grid can be adjusted for
each orbit and each slice, resulting in a considerable speed-up. The
storage-grid must extend a few PSF widths beyond the outermost
observational data points. This PSF convolution is most important for
studying the dynamics at small radii if steep kinematic gradients are
present, \eg in the application of our technique to the search for
massive nuclear black holes.

\subsubsection{Calculating the velocity moments}

After the $k$-th orbit has been calculated, projected onto the storage
cube and convolved with the PSF, we need to extract quantities for
direct comparison with the observational constraints. The data contain
information on the projected properties of the galaxy at a select
number of {\it constraint positions} on the projected face of the
galaxy. Photometry is generally available over the whole face of the
galaxy, extending to much larger radii than the kinematic
measurements. So in general, there are different constraint positions
for the photometric and the kinematic data. In addition, the
constraint positions are often extended areas (\eg the width of a
spectroscopic slit multiplied by the number of pixels along the slit
that were averaged to obtain spectra of sufficient S/N). Clearly, the
storage cube must be chosen sufficiently big to cover all constraint
areas. The constraint positions are labeled by $l$, with
$l=1,\ldots,\Nc$. We denote by $f_{xy,l}$ the fraction of the area of
the storage cube cell centered on the grid point $(x,y)$, that is
contained within the constraint area $l$.

Let $M^k_l$ be the fraction of the total mass on orbit $k$ that
contributes to constraint area $l$. This mass fraction is obtained by
summing over the (PSF convolved) storage cube for the given orbit:
\begin{equation}
  M^k_l= \sum_{xyv} f_{xy,l} \> w^k_{xyv} .
\end{equation}
A dynamical model is determined by its {\it orbital weights}
$\gamma_k$, which measure the fraction of the total mass of the system
that resides on each orbit $k$ (see Appendix~A.1 for details). The
total mass fraction $M_l$ of the model that contributes to constraint
area $l$ is obtained as a sum over all orbits:
\begin{equation}
  M_l = \sum_k \gamma_k \> M^k_l .
\end{equation}
To obtain the observed mass fractions $M^{\rm obs}_l$ at the
constraint positions $l$ from the observed surface brightnesses
$\mu^{\rm obs}_l$, we assume that the {\it stellar population} has the
same mass-to-light ratio everywhere in the galaxy\footnote{
An independently known radial gradient in the stellar
mass-to-light ratio, \eg from a population analysis, 
can be included by scaling the `photometric constraints' at the
beginning of the analysis.}. One then has
\begin{equation}
  M^{\rm obs}_l = \mu^{\rm obs}_l \, A_l / L_{\rm tot}  ,
\end{equation}
where $A_l$ is the area of constraint position $l$, and $L_{\rm tot}$
is the total observed luminosity. Fitting the predicted mass fractions
$M_l$ to the observed mass fractions $M^{\rm obs}_l$ is then a linear
superposition problem for the $\gamma_k$.

For our technique to work, we must also ensure that the contributions
of individual orbits to all kinematic constraints add up linearly.
As we show below, this can be achieved in a straightforward manner
if we choose the Gauss-Hermite coefficients $h_m$~($m=1, \ldots,
M$) to describe the shape of the VP (vdMF; Gerhard 1993). The
normalized VP contributed by orbit $k$ to constraint position $l$ is
\begin{equation}
  {\rm VP}^k_{l,v} = {1\over{M^k_l}} \sum_{xy} f_{xy,l} \> w^k_{xyv} .
\end{equation}
The total normalized VP at constraint position $l$ is obtained as a
sum over all orbits:
\begin{equation}
\label{modelVPs}
  {\rm VP}_{l,v} = {1\over{M_l}} \sum_k \gamma_k \> 
                   M^k_l \> {\rm VP}^k_{l,v} .
\end{equation}
This `histogram', with the velocity $v$ in the subscript on the
left-hand-side as the independent variable, is a discrete
representation of the underlying continuous profile ${\rm VP}_{l}(v)$.
The Gauss-Hermite moment $h_{m,l}$ of order $m$ at constraint position
$l$ is defined as an integral over ${\rm VP}_{l}(v)$:
\begin{equation}
\label{GauHerm}
  h_{m,l} = 2 \sqrt{\pi} \int_{-\infty}^{\infty}
            {\rm VP}_{l}(v) \> \alpha(w_l) \> H_{m}(w_l) \> {\rm d} v.
\end{equation}
The function $\alpha$ is a Gaussian weighting function:
\begin{equation}
  \alpha(w_l) \equiv {1\over{\sqrt{2\pi}}}
                      \exp{\bigl[ -\fr12 w_l^2 \bigr]} .
\end{equation}
The quantity $w_l$ is defined as $w_l \equiv (v-V_l) / \sigma_l$,
where the velocity $V_l$ and dispersion $\sigma_l$ are (for the
moment) free parameters. The $H_m (w_l)$ are Hermite polynomials (see,
\eg Appendix~A of vdMF). One may similarly define the Gauss-Hermite
moment $h^k_{m,l}$ of orbit $k$ and order $m$ for constraint position
$l$, as an integral over ${\rm VP}^k_{l}(v)$ (of which ${\rm
VP}^k_{l,v}$ is the discrete representation). When the free parameters
$V_l$ and $\sigma_l$ are chosen to be the same for each orbit $k$, it
follows that
\begin{equation}
  M_l \, h_{m,l} = \sum_k \gamma_k \> M^k_l \> h^k_{m,l} .
\end{equation}
Thus, fitting the observed Gauss-Hermite moments $h^{\rm obs}_{m,l}$
through the combination $M^{\rm obs}_l \, h^{\rm obs}_{m,l}$ is also a
linear superposition problem for the $\gamma_k$.

In practice we choose $V_l$ and $\sigma_l$ equal to the parameters of
the best-fitting Gaussian to the observed VP at constraint position
$l$ (these are the observationally determined quantities). This
implies $h^{\rm obs}_{1,l} = h^{\rm obs}_{2,l} = 0$ for the first- and
second-order observed Gauss-Hermite moments (vdMF). By requiring the
predicted moments $h_{1,l}$ and $h_{2,l}$ to reproduce this, the model
VP automatically has the correct mean velocity and velocity dispersion
(as determined through a Gaussian fit). Hence, these latter quantities
need not be fitted separately. In this procedure we do require
knowledge of the errors $\Delta h^{\rm obs}_{1,l}$ and $\Delta h^{\rm
obs}_{2,l}$ that correspond to the observationally quoted errors
$\Delta V$ in $V_l$ and $\Delta \sigma$ in $\sigma_l$. These can be
obtained from the general relations for Gauss-Hermite expansions
(vdMF),
\begin{equation}
\label{loworder}
    \Delta h_1 = -{1\over2}\sqrt{2} \> \Delta V / \sigma ,\qquad
    \Delta h_2 = -{1\over2}\sqrt{2} \> \Delta \sigma / \sigma ,
\end{equation}
which are valid to first order in the (small) quantities $(\Delta V /
\sigma)$, $(\Delta \sigma /\sigma)$ and $h_3$, $h_4$, $\ldots$.

The zeroth-order moment $h_0$ defined by equation~(\ref{GauHerm})
measures the normalization of the best-fitting Gaussian to the
normalized VP. This quantity is {\it not} included in the fit, because
it is observationally inaccessible: it is directly proportional to the
unknown difference in line strength between the galaxy spectrum and
the template spectrum used to analyze it. In practice one uses the
assumption $h_0=1$ to estimate the line strength from the
observations. The observational estimates for the higher-order
Gauss-Hermite moments are also influenced by uncertainties in the line
strength, but only to second order. These uncertainties can be safely
ignored in all cases of practical interest.

Our scheme uses the Gauss-Hermite {\it moments} $h_m$, which are
defined as integrals over the VPs. These integrals are well-defined
for arbitrary functions, even highly non-Gaussian ones. Our scheme
therefore assumes neither that the individual orbital VPs are well
described by the lowest order terms of a Gauss-Hermite {\it series}
(which is not generally the case), nor that the observed VPs are well
described by the lowest order terms of such a series (which is
generally the case).

\subsection{Comparison with the observational constraints}

Once the properties of all orbits are calculated for all constraint
positions, we need to find the non-negative superposition of orbital
weights $\gamma_k$ that best matches the observational constraints
within the error bars. When the observational errors are normally
distributed, the quality of the fit to the data is determined by the
$\chi^2$ statistic:
\begin{equation}
\label{chisq}
  \chi^2 \equiv \sum\limits_{l=1}^{\Np} 
            \Biggl ( { { {M^{\rm obs}_l - 
                         {\textstyle \sum} {\gamma_k M^k_l} } \over
                       {\Delta M^{\rm obs}_l}} } 
            \Biggr)^2  + 
        \displaystyle{ \sum_{l=\Np+1}^{\Nc} \, \sum_{m=1}^M
            \Biggl ( { { M^{\rm obs}_l h^{\rm obs}_{m,l}- 
                         {\textstyle \sum} {\gamma_k M^k_l h^k_{m,l}} } \over
                       {\Delta (M^{\rm obs}_l h^{\rm obs}_{m,l})} } \Biggr)^2,
        }
\end{equation}
where we assume that there are $\Np$ photometric and $\Nk = \Nc-\Np$
kinematic constraint positions. Currently, the number $M$ of
Gauss-Hermite moments that can be extracted from spectroscopic
observations is typically 4. In this procedure no arbitrary relative
weighting of the photometry vs.~the kinematics is necessary; the error
on each constraint is objectively taken into account.

When all quantities are divided by their observational uncertainties,
\eg $M^{\rm obs}_1 \rightarrow M^{\rm obs}_1/\Delta M^{\rm obs}_1$,
$M^1_1 h^{1}_{1,1} \rightarrow M^1_1 h^{1}_{1,1} / \Delta (M^{\rm
obs}_1 h^{\rm obs}_{1,1})$, etc., the $\chi^2$ minimization is
converted into a least squares problem:
\begin{equation}
\label{matrixeq}
\left[\matrix{
 M^{1}_1 & \ldots & \ldots & M^{\No}_1\cr
 M^{1}_2&\ldots&\ldots& M^{\No}_2\cr
\vdots&\vdots&\vdots&\vdots&\cr
 M^{1}_{\Np}&\ldots&\ldots& M^{\No}_{\Np}\cr
  M^{1}_{Np+1} h^{1}_{1,\Np+1}&\ldots&\ldots& M^{\No}_{Np+1} h^{\No}_{1,\Np+1}\cr
\vdots&\vdots&\vdots&\vdots&\cr
  M^{1}_{\Nc} h^{1}_{1,\Nc}&\ldots&\ldots& M^{\No}_{\Nc} h^{\No}_{1,\Nc}\cr
\vdots&\vdots&\vdots&\vdots&\cr
\vdots&\vdots&\vdots&\vdots&\cr
  M^{1}_{Np+1} h^{1}_{M,\Np+1}&\ldots&\ldots& M^{\No}_{Np+1} h^{\No}_{M,\Np+1}\cr
\vdots&\vdots&\vdots&\vdots&\cr
  M^{1}_{\Nc} h^{1}_{M,\Nc}&\ldots&\ldots& M^{\No}_{\Nc} h^{\No}_{M,\Nc}\cr
}\right]
\times \left[\matrix{
\gamma_1\cr
\vdots\cr
\vdots\cr
\gamma_{N_o}\cr
}\right] = \left[\matrix{
  M^{\rm obs}_1\cr
  M^{\rm obs}_2\cr
\vdots\cr
  M^{\rm obs}_{\Np}\cr
  M^{\rm obs}_{Np+1} h^{\rm obs}_{1,\Np+1}\cr
\vdots\cr
  M^{\rm obs}_{\Nc} h^{\rm obs}_{1,\Nc}\cr
\vdots\cr
\vdots\cr
  M^{\rm obs}_{Np+1} h^{\rm obs}_{M,\Np+1}\cr
\vdots\cr
  M^{\rm obs}_{\Nc} h^{\rm obs}_{M,\Nc}\cr
}\right],
\end{equation}
which must be solved for the occupation vector
$(\gamma_1,\ldots,\gamma_{\No})$, with the constraints $\gamma_k \ge
0$, for $k=1,\ldots,\No$.  There are standard algorithms for solving
this problem and we use the Non-Negative Least Squares (NNLS)
algorithm by Lawson \& Hanson (1974; see also Pfenniger 1984; Zhao
1996).

The NNLS fit returns the orbital weights
$\gamma_k$ and the model predictions for all
the observed quantities on the right-hand-side of
equation~(\ref{matrixeq}). Among these are the predicted $h_1$ and
$h_2$, but not the predicted $V$ and $\sigma$. In practice it is often
useful to know the latter, for visual comparison to the data. The
predicted $V$ and $\sigma$ can be calculated to first order accuracy
from the predicted $h_1$ and $h_2$ using the
relations~(\ref{loworder}). For higher accuracy one may fit a Gaussian
to the actual VPs predicted by the model, which are obtained by
substituting the $\gamma_k$ into equation~(\ref{modelVPs}).

In practice one considers potentials that depend on a number of
parameters. After the (set of) best-fitting parameter combination(s)
has been determined, the confidence regions on the model parameters
can be estimated from the relative likelihood statistic $\Delta \chi^2
\equiv \chi^2 - \chi^2_{\rm min}$. If the observational errors are
normally distributed, then $\Delta \chi^2$ follows a $\chi^2$
probability distribution, with the number of degrees of freedom equal
to the number of parameters in the potential (Press \etal 1992). Errors on
parameter values quoted in the modeling below correspond to the
$68\%$ confidence level, unless mentioned otherwise.

\subsection{Regularization}

If one uses fewer orbits than constraints, the
NNLS fit will always have a formally unique solution, even when the underlying
physical problem allows a wide range of solutions (an example is the
case in which the observations constrain only the projected mass
distribution and velocity dispersion profile, cf.~Binney \& Mamon
1982).  To produce meaningful results with this method, in practice
one must therefore use more orbits,
or basis vectors of the phase space, than constraints.
In this case, the best solution (which need not be an exact
solution, i.e., $\chi^2=0$) need not be unique. The NNLS fit will
select one of the possible solutions. The adopted solution will
generally be very irregular in phase-space (trying to accommodate all
the noise in the data or the orbit library),
which is physically implausible.
Non-negativity is the only clear-cut physical 
constraint on the distribution
function.
The `smoothness' of the DF in a collisionless system ultimately
depends on the efficiency of the violent relaxation during the
formation of the galaxy, which is difficult to quantify. Here, we are
mostly interested in constraints on the gravitational potential,
independent of the detailed properties of the DF, as long as it is non
negative. Therefore, we employ only a very simple smoothing procedure,
which does not significantly impact the model fit to the data.
%Most further assumptions about the ``smoothness" of the DF
%for a collisionless systems depend on assumptions about the
%efficiency of (violent) relaxation during the formation of the
%galaxy, and are hence all but impossible to specify quantitatively.
%At present, we are mostly interested in statements about the
%gravitational potential, assuming virtually  nothing but non-negativity
%about the DF. Therefore, we will employ only a very simple smoothing
%procedure, and will smooth the DF only to the extent that it has not
%impact on the model fit to the data.

Smoothing of the DF can be  achieved through regularization (Press
\etal 1992), but there is no unique approach (\eg
Merritt 1993b).  Previous authors have either maximized the entropy
of phase space (Richstone \& Tremaine 1988), or have enforced local 
smoothness of the distribution function (Merritt 1993a; Zhao 1996).
Here we use a regularization scheme similar to that of Zhao.  It
minimizes the local curvature of the mass distribution in phase space
by including an additional term to the $\chi^2$-function defined in
equation~(\ref{chisq}):
\begin{equation}
\label{regul}
  \lambda \> \sum\limits_{k=1}^{N_o} \bigl (\hat\gamma_k - {1\over P}
           \sum\limits_{p=1}^P \hat\gamma_{k_p} \bigr ) .
\end{equation}
The $P$ orbits $k_1,\ldots,k_P$ are the `immediate neighbors' of orbit
$k$ in phase space. For our simple $(E_i,L_{ij})$ grid there are four
neighbors to each point that is not on the edge of the grid. The
quantities $\hat\gamma_k$ are defined as $\hat\gamma_k \equiv \gamma_k
/ \gamma^{\rm ref}_k$, where the $\gamma^{\rm ref}_k$ are a set of
reference weights. These could be chosen to reflect any prior knowledge
or prejudice about the DF. For example, they may be set to
the orbital weights that can be calculated semi-analytically
(Appendix~A.1) for an isotropic DF, forcing the model to tend
to the isotropic DF in the limit of infinite smoothing. 

Here we have employed the simplest regularization by
setting all the
$\gamma^{\rm ref}_k$ equal to unity. The parameter $\lambda$ in
equation~(\ref{regul}) governs the degree of regularization. Although
this parameter is in principle freely adjustable, it can be chosen in
a reasonably objective way, by letting the data determine the degree
of permissible DF smoothing.  Let $\chi^2_0$ be the minimum
chi-squared for the case without regularization ($\lambda=0$). Any
solution that matches the data with $\Delta \chi^2 = \chi^2 - \chi^2_0
\lta 1$ is statistically equally acceptable.  We can therefore
increase $\lambda$ until $\Delta \chi^2 = 1$, which yields
a smoother DF that provides an equally good fit
to the data.
This regularization procedure is followed in
all the subsequent applications in this paper.
As Figure 5 shows, this minimal regularization often leads
to drastically smoother DFs with indistinguishable fits to the
data.

Given the simple nature of the regularization employed here, it is
important to reiterate that there is a large class of problems for which
regularization is not essential. For example, this is the case
when the main goal is to rule out certain potentials, \eg those
without a dark halo or without a black hole. If no good fit can be
found without regularization, i.e., allowing arbitrarily un-smooth DFs,
then there will certainly not exist a smooth DF that fits the data.
Thus, by omitting regularization at all, one will always obtain
conservative estimates of the range of potentials that are ruled out.

\section{An illustration of the method}

As an illustration and a test, we create pseudo-data drawn from the
analytically known properties of an isotropic ($f(E)$) non-rotating
Hernquist (1990) model. We assume a mass to light ratio $\Upsilon^{\rm
true}_\star=1$ for the stellar population (in arbitrary units), and
assume that no dark material is present. We match these pseudo-data
with our technique, under the assumption of a mass-to-light ratio
$\Upsilon_\star$. As constraints we use the surface brightness over
the radial range $0.05\,\Reff$ to $2\,\Reff$, with an uncertainty of
5\%, the line-of-sight velocity dispersion with an error of 5\%, and
the (non-zero) values of $h_4$ with an error of $\pm 0.05$ (based on the
observational characteristics of, \eg C95).

We calculated a library of 420 orbits ($N_E=60$ and $N_L=7$), with the
energy grid ranging from $r_{{\rm c},1} = 0.01 \Reff$ to $r_{{\rm
c},N_E} = 6 \Reff$. Each orbit was built up from 25 sub-orbits (see
Section~2.1). Only one orbit library was calculated, for
$\Upsilon_\star=1$. The orbit library for any other $\Upsilon_\star$
is obtained trivially by rescaling the model velocity by a
factor $\sqrt{\Upsilon_{\star}}$. However, the orbit contributions
$h^k_{m,l}$ to the Gauss-Hermite moments at each constraint position
must be calculated separately for each assumed $\Upsilon_{\star}$,
because they involve the observed velocities $V_l$ and dispersions
$\sigma_l$ in a non-linear way. Similarly, the NNLS fit for the
orbital weights $\gamma_k$ must also be done separately for each
$\Upsilon_{\star}$.

We constructed models with $\Upsilon_\star=0.5$, $1$~and $2$. This
mimics the realistic situation in which the true mass-to-light ratio
of a galaxy is unknown, and has to be inferred from models with
different $\Upsilon_\star$. Figure~2 shows the match to the
pseudo-data if only the surface brightness and the velocity dispersion
profiles are fitted.  For the correct mass-to-light ratio (middle
panels), the fit is perfect and the difference between input and
output $h_4$ is negligible, even though $h_4$ was not fitted. Note that 
$\Delta\chi^2=1$ regularization has been applied to all models. The
bottom panel shows the orbital (mass) weights in a modified Lindblad
diagram, where $E$ and $L$ are replaced by $\rc(E)/\Reff$ and
$L/L_{\rm max}(E)$, respectively.  The area of the dots is
proportional to the logarithm of the orbital weight and the vertical
dashed lines show the radial range in which observational constraints
exist. For $\Upsilon_\star=\Upsilon^{\rm true}_\star$ all the mass is
attributed to orbits with $\rc$ within the observed range (although of
course for very radial orbits, \eg $L/L_{\rm max}=0.05$, the apocenter
lies well beyond $\rc$). The model assigns
comparable mass to orbits with the same energy but
different eccentricities.
This is reassuring, because the pseudo-data were
drawn from an isotropic Hernquist model. Note that we have no reason to
expect that our best fitting model is precisely isotropic, because we
are constraining a function of two variables, $f(E, L)$, by two
functions of one variable, the surface brightness and the velocity
dispersion profile.  This is insufficient to fix $f(E, L)$ uniquely,
but the computations show that the remaining freedom in the DF is not
large (see also Dejonghe 1987). We verified that our method does reproduce
the isotropic Hernquist model exactly (to within the small
discretization errors) when only the f(E) components are used (see
Section~2.1 and Appendix~A).
%Because the pseudo-data were drawn from an
%isotropic Hernquist model, they should indeed be matched by an
%isotropic orbit distribution. This was also verified by fitting the
%data with the $f(E)$-components defined in Section~2.1 (and
%Appendix~A), which confirmed that an $f(E)$ model matches the data.

The panels on either side show the fit for potentials with assumed
mass-to-light ratios of $\Upsilon_\star = 0.5$ (left) and
$\Upsilon_\star = 2$ (right). For $\Upsilon_\star = 0.5$, the
resulting distribution function consists of two disjoint pieces: a
tangentially biased part at $\rc \lta \Reff$ and a very radially
biased part at $\rc \gta \Reff$. These parts conspire to increase the
velocity dispersion in the observed range above the isotropic value.
The circular orbits have their highest projected velocity dispersion
at $R \approx \rc$ and the radial orbits have their highest projected
dispersion at radii $R \ll \rc$.  The opposite effect is observed when
$\Upsilon_\star=2$: the orbit weights now are given to radially biased
orbits at small radii and nearly circular orbits at very large
radii. This combination leads to a projected dispersion smaller than
the isotropic value. Despite these discrepancies, it is remarkable
that over much of the radial range the velocity dispersion can be fit
to much better than the factor of $\sqrt{2}$ expected from simple
scaling, owing to the freedom to select a special distribution of
orbits. However, the predicted $h_4$ profiles for both $\Upsilon_\star
= 0.5$ and $\Upsilon_\star = 2$ are significantly non-Gaussian, due to
the awkward phase-space structure; they differ substantially from the
input profile.

Figure~3 shows the same orbit library, but forced to fit to surface
brightness, the velocity dispersion {\it and} the $h_4$ profile. Not
surprisingly, the model with $\Upsilon_\star = 1$ remains virtually
unchanged.  For the other two models the match to $h_4(R)$ is
improved, at the slight expense of the $\sigma (R)$ fit. Figure~4
shows the $\chi^2$ of the model fit as a function of
$\Upsilon_\star$. The dashed line shows the fit if only the surface
brightness and velocity dispersion profiles are used as constraints
(as in Figure~2). A range of potentials, those with $0.8 \lta
\Upsilon_\star \lta 1.2$, all match the data perfectly ($\chi^2
\approx 0$, because no noise was added to the pseudo-data). This is
consistent with the results of Binney \& Mamon (1982); in each
potential the model uses a different orbital structure to fit the
data. The solid line shows the $\chi^2$ when the $h_4$ profile is
included in the fit (as in Figure~3). The range of $\Upsilon_\star$
that produces $\chi^2 \approx 0$ is now much more narrowly centered on
the true value, $\Upsilon_\star^{\rm true}=1$. Many of the potentials
that could fit the surface brightness and velocity dispersion
profiles cannot simultaneously fit the $h_4$ profile.
The VP shape information constrains
the velocity anisotropy, and therefore helps in
limiting the set of allowed potentials.

Figure~5 shows the effect of regularizing the orbital weight
distribution, for the case in which the mass-to-light ratio is
correctly assumed to be $\Upsilon_\star = 1$. The left panel shows the
fit without any regularization: the resulting phase-space distribution
is very jagged, and in fact very different from the isotropic model
that was used to generate the pseudo-data. However, since all orbital
weight distributions that yield fits within $\Delta\chi^2\sim1$ are a
statistically equally good match to the data, we are at liberty to
select the smoothest distribution function amongst those (middle
panel). This model is indeed very close to the isotropic model. The
right panel shows excessive regularization, resulting in
$\Delta\chi^2=10$. In the latter case one sees that the fit to the
data deteriorates if too smooth an orbital weight distribution is
enforced.

\section{Improved mass modeling for elliptical galaxies}

\subsection{Choice of potentials}

The modeling technique yields the relative likelihood of different
gravitational potentials, given the observational
constraints. However, even for the simplest, spherical case, there is
an infinity of trial potentials, $\Phi(r)$. Much of the previous
modeling in the literature has focused on testing the constant
mass-to-light ratio hypothesis, for which the sequence of trial
potentials is one-dimensional, and can be labeled by
$\Upsilon_{\star}$. However, there are two cases of principal interest
in which the total mass density is not proportional to the luminous
stellar density, namely if (a) there is a massive black hole at the
center of a galaxy, or (b) the luminous galaxy is embedded in a dark
halo. While in case (a) the set of trial potentials is characterized
by two parameters, $\Upsilon_{\star}$ and the black hole mass 
$M_{\rm BH}$,
case (b) requires a more complex
treatment. In the application of our technique below, to study the
presence and properties of a dark halo in NGC~2434, we consider three
classes of potentials. In each case the goal is to determine whether
the given class of potentials is consistent with the data, and for
what values of the parameters.

\begin{enumerate}
\item Constant mass-to-light ratio models, representing the case
without a dark halo (or the case in which the dark and luminous matter
have the same spatial distribution). The gravitational potential is
derived directly from the deprojected stellar luminosity (and thus
mass) distribution, with the mass-to-light ratio $\Upsilon_\star$ as
the only free parameter.

\item Logarithmic potentials, as a popular case of an ad hoc
functional form for the gravitational potential of a galaxy with a
dark halo, with the (constant) circular velocity $\Vc$ as the
only free parameter.

\item Cosmologically motivated `star+halo' models, based on the recent
work by NFW. These models use dark matter mass profiles predicted from
collisionless cosmological simulations, which are modified by the
baryonic/stellar mass accumulating at their center under the
assumption of adiabatic invariance. The motivation and construction of
these potentials is described in detail in Appendix~B. The resulting
potentials are characterized by two parameters: the stellar
mass-to-light ratio $\Upsilon_{\star}$ and a characteristic scale
velocity $V_{200}$ of the dark halo. 
\end{enumerate}

\subsection{An application: The dark matter halo around NGC~2434}

We combine the technique discussed in Section~2 with the sequence of
trial potentials described in Section~3.1, to ask what range of
gravitational potentials are compatible with the observed photometry
and kinematics of NGC~2434. This is a nearly round (E0) elliptical
galaxy at 20 $\Mpc$, with absolute luminosity $M_B=-19.9$ ($1.4 \times
10^{10}L_\odot$; $H_0=70 \kms \Mpc^{-1}$ throughout this paper).
Rotation is unimportant ($E_{\rm rot}/E_{\rm kin}\sim 0.01$).  The
kinematic data ($\sigma$ and $h_4$) are from C95, and extend to
$60\asec$; the photometry is from Carollo \& Danziger (1994), and
extends to $105\asec$.  NGC 2434 is one of the few early-type galaxies
where the stellar kinematics, including the shape of the VP, have been
measured to $\sim 2.5 \Reff$ ($\Reff\sim 24\asec$).

C95 showed that the kinematics of NGC 2434 could not be fit with
axisymmetric constant mass-to-light ratio models with a DF of the form
$f = f(E,L_z)$ (for the case of NGC 2434, which is nearly round on the
sky, these reduce essentially to spherical isotropic $f(E)$
models). From the sign of the discrepancy between their simple models
and the data, they also {\it inferred} that no other model without dark
matter would fit the observational constraints.  Here we take the
analysis two steps further: (i) we {\it demonstrate} that NGC~2434's
kinematics cannot be fit by any constant mass-to-light ratio model,
regardless of the anisotropy of the orbital distribution; and (ii) we
explore the issue of {\it how much} dark matter is required to match
the data.

\subsubsection{Constant mass-to-light ratio models}

We parameterized the luminosity density as
\begin{equation}
  \rho_\star \propto (r/b)^{\alpha}~\bigl [1+(r/b)^2\bigr ]^{\beta}, 
\end{equation}
(with $\alpha=-1.9$, $\beta=-0.55$ and $b=13\asec$, see C95), and
calculated the luminous gravitational potential upon assumption of a
constant mass-to-light ratio $\Upsilon_\star$.

As in the test case of Section~2.5, an orbit library was calculated
for only one value of $\Upsilon_\star$, which was subsequently scaled
to arbitrary values of $\Upsilon_\star$. We used a grid of $60 \times
8$ orbits in $E$ and $L/L_{\rm max}$, with each orbit `built' from 25
sub-orbits. The energy grid ranged from $r_{{\rm c},1} = 0.1\asec$ to
$r_{{\rm c},N_E} = 500\asec$. Our technique was used to fit the data
(with regularization as discussed in Section~2.4), for B-band
mass-to-light ratios $\Upsilon_{\star,B}$ ranging from 3.75 to 10 (in
solar units).

Figure~6 shows the fits for various values of $\Upsilon_{\star,B}$.
The best fit is found for $\Upsilon_{\star,B} = 6.2 \pm 0.3$, but the
observational data cannot be well fit by these models for {\it any}
value of $\Upsilon_{\star,B}$. For all values the algorithm invokes a
highly anisotropic DF, in which the anisotropy is radically different
between the inner and outer parts of the galaxy. However, at large
radii none of the models can match a dispersion profile as flat as
observed, while maintaining the observed nearly Gaussian VPs. Compared
to the best-fitting models presented in Section~3.2.3, the constant
mass-to-light ratio models can be ruled out at the $>99.9$\%
confidence level, independent of the orbital anisotropy of the system.

It is worth noting that this rejection of the constant mass-to-light
ratio hypothesis requires the observed constraints on the VP,
i.e., the knowledge that the VP is approximately Gaussian $|h_4|<0.1$.
If only the surface brightness and the velocity dispersion are used
as constraints, an (almost) acceptable model can be found with
constant mass-to-light ratio.

\subsubsection{Logarithmic potentials}

Scale-free logarithmic potentials have $\Phi(r) \equiv \Vc^2~\log{r}$.
Such potentials have a constant circular velocity $\Vc$, and have
therefore been popular as approximations to the potentials of galaxies
with dark halos. Although the stellar kinematics of elliptical
galaxies at $R \lta \Reff$ can be well fit by models with logarithmic
potentials (Kochanek 1994), constant mass-to-light ratio models
generally provide equally good fits (van der Marel 1991). However,
studies of gravitational lensing statistics (Maoz \& Rix 1993) and
gravitational lensing observations for individual galaxies (\eg
Kochanek 1995) strongly rule out these constant mass-to-light ratio
models, whereas logarithmic potentials do provide good
fits. Logarithmic potentials therefore provide the logical next step
in our modeling of NGC 2434.
 
We calculated orbit libraries for logarithmic potentials with $\Vc$
ranging from 250 to $370 \kms$, in similar fashion as in
Section~3.2.1. A model with $\Vc = 300 \pm 15 \kms$ was found to
provide an excellent fit to the data, as shown in Figure 7. The
distribution function is smooth and contiguous, and close to
isotropic. The small error bar on $\Vc$ illustrates that the addition
of VP shape constraints allows the normalization of the potential to
be accurately determined, provided that its shape is assumed to be
known a priori.

\subsubsection{Cosmologically motivated star+halo potentials}

So far, we have shown that constant mass-to-light ratio models fail,
whereas a model with a logarithmic potential succeeds in fitting the data
for NGC 2434. The cosmologically motivated star+halo potentials
discussed in Appendix~B provide a continuous sequence that connect
these two cases. We computed a grid of these potentials, with
$\Upsilon_{\star,B}$ in the range $1.5$--10 and $V_{200}$ in the range
of 0--$800 \kms$ ($V_{200}=0 \kms$ corresponds to the constant
mass-to-light ratio model). Circular velocity curves for some of the
resulting potentials are shown in Figure~8. We calculated orbit
libraries in these potentials, in similar fashion as in Section~3.2.1.
For each $(\Upsilon_{\star,B}, V_{200})$ the orbital distribution was
found that best reproduced the data, with the corresponding $\chi^2$.
With these potentials an excellent fit can be obtained to the data. As
an example, Figure~9 shows a model with $\Upsilon_{\star,B} = 3.5$ and
$V_{200}=400 \kms$.

Figure~10 shows the grid of models that were calculated to explore
the relative likelihood of models in the 
$(\Upsilon_{\star,B}, V_{200})$ plane.
The area of each point corresponds to the logarithm of their relative
likelihood. There is a
clear anti-correlation between $\Upsilon_{\star,B}$ and the mass of
the halo (proportional to $V_{200}^2$). This is because the most
robustly constrained quantity is the mass inside a characteristic
radius ($\sim R_{\rm eff}$), which could be either stellar or dark.
This anti-correlation is quantified in Figure 11, which shows the 68\%
and 95\% confidence regions for the joint distribution
of $(\Upsilon_{\star,B}, V_{200})$.
Figure 11 also shows that the best fitting parameters are
$\Upsilon_{\star,B}=3.35\pm0.25$ and $V_{200} = 450 \pm 100 \kms$.
The errors quoted refer to the 68\% confidence region of each
parameter individually.

Interestingly, the stellar mass-to-light ratio $\Upsilon_{\star,B}$
for the best fit models is only 50\% of the value for the best-fitting
constant mass-to-light ratio model shown in Figure~6, which is,
however, a poor fit. Even with a minimum
halo of $V_{200} = 250 \kms$, roughly the smallest value allowed by
the data, it is still only 62\%. Thus constant mass-to-light ratio
models tend to overestimate the true stellar mass-to-light ratio when
a dark halo is present. The reason is that these models attempt to
reproduce roughly the correct mass within a characteristic radius
($\sim R_{\rm eff}$), but have no other option than to ascribe this
mass to the stars.

The best-fitting models (e.g., that in Figure~10) have circular
velocity curves that are nearly flat (see Figure~8). Their potential
resembles that of the best-fitting logarithmic potential in Figure~9.
As a result, their orbital distribution is also very similar.

\section{Conclusions}

We have described an extension of Schwarzschild's method that is
capable of modeling the full line-of-sight velocity distributions of
galaxies. Similarly to what is done in the original scheme, we build
galaxy models by computing a representative library of orbits in a
given potential, and determine the non-negative superposition of these
orbits that best fits an observed set of photometric and kinematic
constraints.  We have extended the technique to predict the full VP
shapes for all models, and to include them in the fit to
the data as a set of linear constraints. This allows us to fully
exploit the high quality kinematical and VP shape data that are now
becoming available, and so to constrain the anisotropy of the 
velocity distribution.  The latter has
been the main uncertainty in previous attempts to infer the
gravitational potentials of elliptical galaxies from observational
data. Explicit modeling of VPs also removes systematic errors in the
comparison to the lowest-order velocity moments $V$ and $\sigma$.

We characterize the VPs through their Gauss-Hermite moments.
Measurements of observed VPs are now routinely specified
by constraints on these moments.
This is useful in practice, because data on
observed VPs are often reported as constraints on the Gauss-Hermite
moments. It is also computationally convenient, because it reduces VP
histograms with 30--50 bins to a small number of
parameters. However, our technique is not restricted to the use of
Gauss-Hermite moments. In principle one may fit individual velocity
bins of observed VPs, if such data is available (as it is, e.g., in
the technique of Rix \& White 1992).  However, this does increase
storage and memory requirements, while also increasing the size, and
therefore the computational complexity, of the NNLS fit.

We also extend Schwarzschild's method as a modeling tool by
taking into account the error on each
observational constraint in finding the best matching orbit
superposition. Hence we
obtain an objective measure for the quality-of-fit. Only projected,
observable quantities are included in the fit. 
We have also outlined how to include the PSF and the detailed
sampling of the data in the data--model comparison by convolving each
orbit and sampling it over the observational apertures. 
This procedure is especially important in the study of galactic nuclei. 
We enforce
smoothness of the model DFs in phase space through a simple
regularization scheme.

The scheme presented in Section 2 is valid for any geometry.
However, in this first paper, we restricted ourselves to spherical
systems which allows swift numerical calculation of the orbit
libraries. The orbit superposition scheme was presented in its
general form.  This allows us to test new
elements of our technique in the most straightforward way.
We tested our method by constructing
models that reproduce pseudo-data drawn
from an isotropic Hernquist model, with and without inclusion of VP
shape constraints in the fit. The results clearly demonstrate the
importance of these constraints in narrowing down the range of
potentials that can fit a given velocity dispersion profile
when no assumptions are made about the
form of the DF.

As an application we have considered the case of dark halos around
elliptical galaxies. We used our technique to interpret the stellar
kinematical and VP shape data out to $\sim 2.5 \Reff$ for the E0
galaxy NGC~2434. Models were constructed with a constant mass-to-light
ratio, with logarithmic potentials, and with cosmologically motivated
star+halo potentials. The latter are based on the cosmological
simulations by NFW, but are modified to incorporate the accumulation
of baryonic matter under the assumption of adiabatic
invariance. Models without a dark halo are ruled out. Both a
logarithmic potential and a cosmologically motivated star+halo
potential can provide an excellent fit to the data. The dark halo of
NGC 2434 is such that roughly half of the mass within an effective
radius is dark. Models without a dark halo therefore tend to
overestimate the mass-to-light ratio of the stellar population by a
factor of $\sim 2$. The best-fitting star+halo potential has a
circular velocity curve that is constant (`flat') to within $\sim 10\%$
from $0.2 \Reff$ to $3\Reff$. This constant circular velocity is close
to that of the best-fitting logarithmic potential, which has $\Vc =
300 \pm 15 \kms$.

The results presented here provide a first test of the conjecture
that elliptical galaxies and spirals of the same stellar mass started out
in similar dark matter halos. Obviously, the two types of galaxies differ
in the degree to which the baryons were concentrated at their centers
(with ellipticals being much denser). If the `baryon contraction' led
to a flat rotation curve in spirals, it should have led to a centrally
peaked (and hence outward falling) rotation curve in ellipticals.
The nearly constant circular velocity in NGC 2434 is, however,
inconsistent with this idea.

The extension to axisymmetric systems is described in detail in a
forthcoming paper (Cretton \etal 1997). The application to the galaxy
M32, to investigate the presence of a massive central black hole, is
discussed in van der Marel \etal (1997a,b). A further extension to
tumbling triaxial systems is in progress. Another possible extension
is to include radial variations in the stellar population. This can be
achieved by choosing a radially varying $\Upsilon_\star$, or by
including the color and line-strength gradients in the set of
constraints, while allowing each orbit to be occupied by stars with
different physical properties.

\acknowledgments 

It is a pleasure to thank Martin Schwarzschild for stimulating
conversations, and HongSheng Zhao for a critical reading of the
manuscript. NC acknowledges financial support from NUFFIC and from the
Leids Kerkhoven Bosscha Fonds, and the hospitality of Steward
Observatory. CMC and RPvdM were supported by NASA through Hubble
Fellowships, \#HF-1079.01-96A and \#HF-1065.01-94A,
 respectively, awarded by the
Space Telescope Science Institute which is operated by the Association
of Universities for Research in Astronomy, Incorporated, under NASA
contract NAS5-26555.

\clearpage
\appendix

\section{The differential mass density}

We describe here the connection between the orbital weights used in
the Schwarzschild technique and the DF. Our treatment largely follows
that of Vandervoort (1984).

\subsection{Inferring the distribution function from the orbital weights}

The general DF for a spherical system is $f(E,L)$, where the binding
energy $E$ and angular momentum $L$ per unit mass are
\begin{equation}
  E = \Psi(r) - {1\over 2} v_r^2 - {L^2 \over 2r^2}, \qquad
  L = r \vt,
\end{equation}
and the positive gravitational potential is $\Psi \equiv -\Phi$.  A
solution of the Schwarzschild technique is not a direct approximation
to $f(E,L)$, but to the differential mass density in the Lindblad
diagram, $\gamma(E,L) \equiv {\rm d}M \> / \> {\rm d}E \, {\rm d}L$.
This function is fully determined by the DF and the gravitational
potential.

The total mass of the system is the integral of the DF over
phase-space
\begin{equation}
  M = \int {\rm d}^3 {\bf r} \, {\rm d}^3 {\bf v} \, f(E,L) \\
    = \int 4\pi r^2 \, {\rm d}r  
      \int \, {\rm d}\xi 
      \int \vt \, {\rm d}\vt
      \int {\rm d}v_r \, f(E,L) ,
\end{equation}
where the angle $\xi$ is the direction of the tangential velocity
vector, as defined in Section~2.2.1. The integral over $\xi$ is trivial
(since neither $E$ nor $L$ depends on it), so $M$ results from a
three-dimensional integral over the phase-space coordinates
$(r,\vt,v_r)$. One may change the order of the integrations and change
to the integration variables $(r,L,E)$ to obtain
\begin{equation}
\label{Mdef}
   M = 8 \pi^2 
         \int_{-\infty}^{\Psi(0)} {\rm d}E
         \int_0^{L_{\rm max}(E)} L \, f(E,L) \, T_r (E,L) \, {\rm d}L .
\end{equation}
The radial period $T_r(E,L)$ of the orbit with integrals $(E,L)$ 
is defined as
\begin{equation}
   T_r (E,L) \> \equiv \> 
       2 \int_{r_{\rm min}(E,L)}^{r_{\rm max}(E,L)} 
           { {{\rm d}r} \over {v_r} } 
             \> = \>
       2 \int_{r_{\rm min}(E,L)}^{r_{\rm max}(E,L)} 
           { {{\rm d}r} \over 
             { [2(\Psi(r)-E) - (L^2/r^2)]^{1/2} } } ,
\end{equation}
and can be easily calculated numerically for any
potential. Equation~(\ref{Mdef}) shows that the differential mass
density is:
\begin{equation}
  \gamma(E,L) = 8 \pi^2 \, L \, f(E,L) \, T_r (E,L) .
\end{equation}

The Schwarzschild technique characterizes a system by the orbital
weights (mass fractions) $\gamma_k$ for a discrete set of orbits. Each
orbit $k$ corresponds to a combination $(E_i,L_{ij})$ of the integrals
of motion.  Let the corresponding grid cell in $(E,L)$ space have an
area $A_{ij}$. The orbital weights $\gamma_k$ are then related to the
DF according to
\begin{equation}
\label{gamDF}
  \gamma_k \approx 8 \pi^2 \, 
      L_{ij} \, f(E_i,L_{ij}) \, T_r (E_i,L_{ij}) \, A_{ij} / M .
\end{equation}
This relation allows the DF to be estimated from a set of $\gamma_k$
inferred by the technique (or, it allows the $\gamma_k$ to be
predicted for a test model with a known DF).

\subsection{Construction of $f(E)$-components}

It proved useful for testing our technique to use building blocks that
are more complex than individual orbits. In particular, we constructed
building blocks that correspond to a weighted sum of orbits with the
same energy. The orbital occupancies $W_{xyv}$ of these components
are related to the orbital occupancies $w_{xyv}$ for the individual
orbits, defined in Section~2.2.1, through:
\begin{equation}
\label{Wdef}
   W_{xyv}^{i} = \sum_{j=1}^{N_L} \epsilon_{j} \, w_{xyv}^{ij} 
\end{equation}
(we denote here each orbit by the indices $i$ and $j$ of its position
on the $(E,L)$ grid, rather than by its index $k$). The weights
$\epsilon_{j}$ are such that each component corresponds to
an isotropic DF, restricted to one energy $E_i$. This requires that 
\begin{equation}
\label{epsdef}
   \epsilon_{j} = L_{ij} \, T_r (E_i,L_{ij}) \, A_{ij} \> \big / \>
                   \left [
                      \sum_{j=1}^{N_L} L_{ij} \, T_r (E_i,L_{ij}) \, A_{ij} 
                   \right ] .
\end{equation}
This follows from equation~(\ref{gamDF}) (which gives the relative
masses contained on orbits with the same energy but different angular
momenta) and from the fact that the $\epsilon_j$ must add to unity (so
that the orbital occupancies $W^i_{xyv}$ are normalized). We call the
resulting building blocks `$f(E)$-components'. Any combination of
these components yields an isotropic DF.

%Equations~(\ref{Wdef}) and~(\ref{epsdef}) can also be derived from an
%analysis based on delta-function DFs. Equation~(\ref{Mdef}) gives the
%total mass of a component
%that is a delta-function in the energy, $f = A_i \>
%\delta(E-E_i)$, or of an orbit that is a delta-function in the both energy
%and angular momentum, $f = B_{ij} \> \delta(E-E_i,L-L_j)$. DFs that are
%normalized to unit mass require 
%%
%\begin{equation}
%  A_i = \Biggl [8\pi^2 \, \int T_r(E_i,L) L \, {\rm d} L \Biggr]^{-1} , \qquad
%  B_{ij} = \Biggl [8\pi^2 T_r(E_i,L_{ij}) L_{ij} \Biggr]^{-1} .
%\end{equation}
%%
%The projected properties for such DFs follow from the previous
%expressions. For instance, the VP of an orbit that has $E = E_i$ and
%$L = L_i$ can be evaluated by a 1D quadrature over the
%line of sight
%%
%\begin{eqnarray}
%  \VP(x,y,v) &=& \int f(E_i,L_{ij}) \; \d v_x \, \d v_y \, \d z \nonumber \\
%             &=& \int f(E_i,L_{ij}) \,{\rm J} \; \d E \, \d L \, \d z \\
%             &=& B_{ij} \int {\rm J} \;\d z \nonumber
%\end{eqnarray}
%%
%where $J$ is the Jacobian for the change of integration variables, from 
%$(v_x,v_y)$ to $(E,L)$:
%%
%\begin{equation}
%  {\rm J} = \Biggl[2 v_y x (x v_x + z v)\Biggr] ^{-1}
%\end{equation}
%%
%with 
%%
%\begin{eqnarray}
%v_y &=& \pm \sqrt{2(\Psi - E_i) - v^2 - v_x^2} \nonumber \\
%v_x &=& -{1 \over x}\Bigl[z v \pm \sqrt{2(\Psi - E_i)(z^2 + x^2) - L_i^2}\Bigr] 
%\end{eqnarray}
%%%%%%%%%%%%%%%%%%%%%%%%%%%%%%%%%%%%%%%%%%%%%%%%%%%%%%%%%%%%%%%%%%%%%%

One may alternatively derive the construction of the $f(E)$-components
from an analysis based on delta-function DFs. Let $f^{\delta}_{[E_0]}$
represent a delta-function in the energy, $f^{\delta}_{[E_0]}(E)
\equiv A_0 \> \delta(E-E_0)$, and let $f^{\delta}_{[E_0,L_0]}$
represent a delta-function in both energy and angular momentum,
$f^{\delta}_{[E_0,L_0]}(E,L) \equiv B_{0} \> \delta(E-E_0,L-L_0)$.
These DFs are normalized to unit mass if
\begin{equation}
  A_0 = \Biggl [8\pi^2 \, \int L \, T_r(E_0,L) \, {\rm d} L \Biggr]^{-1} , 
     \qquad
  B_0 = \Biggl [8\pi^2 L_0 \, T_r(E_0,L_0) \Biggr]^{-1} ,
\end{equation}
cf.~equation~(\ref{Mdef}). Therefore, 
\begin{equation}
  f^{\delta}_{[E_0]}(E) =
      \int f^{\delta}_{[E_0,L_0]}(E,L)
              {L_0 \, T_r(E_0,L_0) \over 
               \int L' \, T_r(E_0,L') \, \d L'} \; \d L_0 ,
\end{equation}
which shows how $f^{\delta}_{[E_0]}$ is build up as a weighted
integral over the $f^{\delta}_{[E_0,L_0]}$. The weights provide the
continuous analog of equation~(\ref{epsdef}). The projected properties
for the delta-function DF $f^{\delta}_{[E_0,L_0]}$ can be traced in
analytical fashion (see also Merritt 1993a). For example, the VP can
be evaluated through a 1D quadrature over the line of sight:
\begin{eqnarray}
  \VP(x,y,v) &=& \int f^{\delta}_{[E_0,L_0]} (E,L) 
                      \> \d v_x \, \d v_y \, \d z \nonumber \\ 
             &=& \int f^{\delta}_{[E_0,L_0]} (E,L)
                      \, J(E,L) \> \d E \, \d L \, \d z 
             \> = \> B_0 \int J(E_0,L_0) \> \d z ,
\end{eqnarray}
where $J(E,L)$ is the Jacobian for the change of integration variables, from 
$(v_x,v_y)$ to $(E,L)$:
\begin{equation}
  J = \Biggl[2 v_y x (x v_x + z v)\Biggr] ^{-1} ,
\end{equation}
with 
\begin{equation}
v_y = \pm \sqrt{2(\Psi - E) - v^2 - v_x^2} , \qquad
v_x = -{1 \over x}\Bigl[z v \pm \sqrt{2(\Psi - E)(z^2 + x^2) - L^2}\Bigr] .
\end{equation}
%%%%%%%%%%%%%%%%%%%%%%%%%%%%%%%%%%%%%%%%%%%%%%%%%%%%%%%%%%%%%%%%%%%%%%

%Finally isotropic building blocks $f(E_i)$ can be constructed as 
%a weighted sum of all the orbits of the same energy $E_i$:
%%
%\begin{equation}
%  f(E_i) = \int f(E_i,L_i) \cdot {T_r(E_i,L)\,L \over 
%                  \int T_r(E_i,L) \; L\,\d L} \; \d L.
%\end{equation}

\section{Cosmologically motivated star+halo models}

To model the dynamics of galaxies with dark matter halos one must
choose a dark potential to add to the luminous one. The traditional
approach, used widely when fitting rotation curves of spiral galaxies,
has been to describe the dark matter as an isothermal sphere with
asymptotic circular velocity $\Vc$ and core radius $\rh$ (\eg van
Albada \etal 1985). The main drawback of this approach is that it is
without physical basis: it is implausible to expect a dark matter halo
with a constant density core after the baryonic mass has condensed and
concentrated at its center. It also has disadvantages from a practical
point of view. The dark halo is described by two new parameters, which
add to the unknown mass-to-light ratio $\Upsilon_{\star}$ of the
stellar population. This makes the sequence of potentials that must be
compared to the data three-dimensional, which makes a proper
comparison to data for elliptical galaxies very time consuming. Apart
from this, $\Vc$ and $\rh$ are highly correlated and cannot even be
determined independently from observed rotation curves for most spiral
galaxies.

Suggestive alternatives come from cosmological studies. NFW and Cole
\& Lacey (1996) have found that in numerical simulations of many
cosmogonies, including standard CDM, the spherically averaged density
profiles of the forming virialized halos can be described by a simple
functional form over a wide range of radii:
\begin{equation}
\label{NFWprof}
  \rho(r)={{\rho_{\rm crit}~\delta_c}
            \over{ (r/r_s)\bigl [1+(r/r_s)\bigr]^2 } } ,
\end{equation}
where $\rho_{\rm crit} = (3H^2_0)/(8\pi G)$ is the critical density of
the universe. The two scale parameters $\delta_c$ and $r_s$ are defined
in terms of a dimensionless concentration parameter $c$, through
\begin{equation}
  \delta_c={{200}\over{3}}~c^3 / \Bigl[ \ln{(1+c)} - c/(1+c)\Bigr] , \qquad
  r_s = r_{200}/c  ,
\end{equation}
where $r_{200}$ is defined as the radius within which the mean halo
density is $200\rho_{\rm crit}$. Each halo may also be described by a
mass scale $M_{200}\equiv 200\rho_{\rm crit} (4\pi/3)r^3_{200}$ and a
velocity $V_{200}\equiv \sqrt{ GM_{200}/r_{200} }$. Mathematically,
any two of these parameters are sufficient to describe the halo's
structure, \eg $V_{200}$ and $c$. Remarkably, NFW found
in addition that for a given cosmology, the two parameters are 
tightly correlated (\eg with $\sim 15\%$ scatter in $c$ at a given
$V_{200}$). For standard CDM this relation can be described by
\begin{equation}
  \log_{10} \, c = 1.05 \> - \> 0.15
     \bigl [M_{200}/ (3 \times 10^{13} M_{\odot}) \bigr ] .
\end{equation}
Thus, these N-body simulations essentially suggest a cosmologically
motivated, one-parameter sequence of dark matter halo models.

The NFW simulations do not contain dissipative, baryonic matter. In
reality, such matter collects at the center of the potential well to
form the visible part of the galaxy. Blumenthal \etal (1986) suggested
that adiabatic invariance can be exploited to estimate how the halo
structure is modified by the accumulation of luminous matter at the
center. This was confirmed by N-body simulations with a crude model
for the infall of the dissipative matter. In general, adiabatic
invariance holds (approximately) for those orbits with periods smaller
than the characteristic time scale of the changes in the potential
(Binney \& Tremaine 1987). An elliptical galaxy has a dynamical time
of a few times $10^7$ years at the effective radius, so adiabatic
invariance may be a good approximation if the baryonic infall occurs
over a time scale exceeding $10^{7-8}$ years.

If the radial mass profiles of baryons and dark matter were initially
the same, then adiabatic invariance in the simplest approximation
(Blumenthal \etal 1986) relates the initial and final radii of the
mass shells $r_{\rm i}$ and $r_{\rm f}$, respectively, by
\begin{equation}
  r_{\rm f} \Bigl [ M_\star(r_{\rm f}) + M_{\rm DM}(r_{\rm f}) 
            \Bigr ] = r_{\rm i} M_{\rm i}(r_{\rm i}) .
\end{equation}
Here $M_{\rm i}(r_{\rm i})$ is the initial total mass within radius
$r_{\rm i}$, $M_\star(r_{\rm f})$ is the final stellar mass within
radius $r_{\rm f}$, and $M_{\rm DM}(r_{\rm f})$ is the final dark mass
with radius $r_{\rm f}$. We assume the initial profile $M_{\rm
i}(r_{\rm i})$ to be as given by NFW for the case of standard CDM
($\Omega=1$), with the characteristic velocity $V_{200}$ as free
parameter. We then employ adiabatic contraction to obtain the final
dark matter profile $M_{\rm DM}(r_{\rm f})$ from the observed stellar
mass profile $M_\star(r_{\rm f})$, assuming a stellar mass-to-light
ratio $\Upsilon_{\star}$. This yields a sequence of star+halo models
that are characterized by the parameters $\Upsilon_{\star}$ and
$V_{200}$. This procedure does not account for mergers, which may
decrease the density of the luminous and dark matter. In this sense
the present procedure, assuming $\Omega=1$ and neglecting late
mergers, leads to the densest possible halos, at a given $v_{200}$.

Circular velocity curves for some of the resulting potentials are
shown in Figure~8. Several features are noteworthy: (i) the transition
from the luminous to the dark matter dominated regime is always
smooth, and is not marked by a feature in the circular velocity curve;
(ii) because the condensing baryons drag some of the dark matter with
them to the center, the dark matter fraction decreases slower towards
small radii than it does for constant density core halos; and (iii)
compact halos are required in order to produce an `effectively flat'
circular velocity curve.

{}

\clearpage

\noindent{ {\bf Figure 1~} Grey-scale representations of the occupation weights
$w^k_{xyv}$ of four orbits, shown in the $(R,v)$ plane of projected
radius vs.~line-of-sight velocity. The orbits have the same energy
$E$, but different angular momenta $L/L_{\rm max}$. Each orbit is assembled
from 25 sub-orbits, as described in \S 2.1.
The axes are in units of the cell sizes, described in \S 2.2.1.
The orbits were calculated in a Hernquist
potential at an energy for which $r_c(E) \approx \Reff$. Note the
changes in the `caustic' structure from the nearly radial orbit at the
top to the nearly circular orbit at the bottom, reflecting the changes
in peri- and apocenter distance.}

\noindent{{\bf Figure 2~}
Fits to pseudo-data (small dots with error bars) drawn from a
spherical, isotropic, non-rotating Hernquist model with $\Upsilon^{\rm
true}_\star=1$. In each column we show (from top to bottom) surface
brightness $\mu$ (in magnitude units), velocity dispersion $\sigma$
(arbitrary units) and the VP shape coefficient $h_4$, all as a
function of radius. The solid lines in each panel (in some cases
coinciding with the sequence of dots), show the model prediction when
only $\mu(R)$ and $\sigma(R)$, {\it but not} $h_4(R)$ were
fitted. The bottom panel in each column shows the orbital weights of
the best fit model in the Lindblad diagram. The area of each dot is
proportional to the logarithm of the orbital weight. The axes of the
Lindblad diagram are not the conventional $(E,L)$, but rather
$\rc(E)/\Reff$ on the abscissa and $L/L_{\rm max}(E)$ on the
ordinate. For the model in the center column the orbits were
calculated in the potential with the correct value of
$\Upsilon_\star$. The panels on the left and right show the model
predictions if $\Upsilon_\star$ is assumed to be $0.5$ and $2$,
respectively.}

\noindent{{\bf Figure 3~}
As Figure~2, but now all observational constraints,
$\mu(R)$, $\sigma(R)$ {\it and} $h_4(R)$ were fitted.}

\noindent{{\bf Figure 4~}
Illustration of the additional constraints on the
gravitational potential provided by VP shape information. The dashed
line shows $\chi^2$ for the fit to the Hernquist test model with
$\Upsilon_{\star}^{\rm true}=1$, when only $\mu(R)$ and $\sigma(R)$
are fitted. A range of potentials, $0.8<\Upsilon_\star<1.2$, provides
a perfect fit to the pseudo-data ($\chi^2 \approx 0$, because no noise
was added).  The assumption of an incorrect mass-to-light ratio can be
fully compensated by a change in the orbital distribution. This is no
longer true if the model must also fit $h_4(R)$ (solid curve). In this
case the range of potentials that provides a perfect fit is much more
narrowly centered on the true mass-to-light ratio.}

\noindent{{\bf Figure 5~}
The effect of regularization on the modeling. All models in
this figure have the correct mass-to-light ratio. The left panels show
the best fit without any regularization. The phase space distribution
is jagged in response to the numerical noise in the projected orbit
properties. However, all models that produce a $\chi^2$ that does not
differ by more than $\Delta\chi^2 = 1$ are statistically equally
acceptable. Hence, we are at liberty to select the smoothest of these
models (center column; same as the center-column of Figure~3). The
right column shows that with excessive regularization the DF is very
smooth, but at the expense of a worse fit to the data.}

\noindent{{\bf Figure 6~}
Predictions of constant mass-to-light ratio models compared
to the data for NGC 2434. The top panels show $\Delta (\mu-\mu_{1/4})$
the difference between the surface brightness (in magnitudes)
and the surface brightness of the best fitting de~Vaucouleurs model.
All other quantities in the panels are as in
Figure~2. Models are shown for, from left to right,
$\Upsilon_{\star,B} = 5$, $6.2$ and~$8.7$. The middle column provides
the best fit, but the observational data cannot be well fit for {\it
any} value of $\Upsilon_{\star,B}$. This implies that constant
mass-to-light ratio models are ruled out, independent of the orbital
anisotropy of the system.}

\noindent{{\bf Figure 7~}
Predictions of a model with a logarithmic potential with $\Vc
= 300 \kms$, compared to the data for NGC 2434. The model provides an
excellent fit, by contrast to the constant mass-to-light ratio models
in Figure~6. All observational constraints are matched within the
error-bars with a fairly smooth and contiguous distribution of
orbital weights. The value of $\chi^2=25$ is considerably lower than
expected for the fit to 98 data points. It is apparent from the
Figure that this is not because the model ``fits every noise spike",
but because a portion of the errors in the photometry and kinematics
is systematic; i.e. the scatter among neighboring data points
is less than the size of their error bars.}

\noindent{{\bf Figure 8~}
Rotation curves, $v_c=(R\partial\Phi/\partial R)^{1/2}$, for
four of the cosmologically motivated star+halo potentials described
in Appendix~B (solid curves). In all cases the stellar mass-to-light
ratio $\Upsilon_{\star,B} = 3.5$, but the characteristic halo velocity,
$V_{200}$, changes from 100 to $600 \kms$. The dashed and dash-dotted
curves show the separate contributions from the stars and the dark
matter, respectively. The sequence presents a way of
consistently and continuously `bending up' the outer part of the
rotation curve. The mass distribution for $V_{200}=400 \kms$
leads to an approximately flat rotation curve from $0.2 < r/\Reff < 3$ and is
very similar to the best fitting logarithmic potential (Figure~7).}

\noindent{{\bf Figure 9~}
Predictions of a model with a cosmologically motivated
star+halo potential, compared to the data for NGC 2434. The fit is
excellent. The model has $\Upsilon_{\star,B}=3.5$ and $V_{200} = 400
\kms$, and was found to provide the best fit among the parameter combinations
studied, cf.~Figure 10. The fit to the data and the orbital
structure of this model are very similar to those in Figure~7 for the
logarithmic potential, as might be expected from the similarity in the
two potentials (cf.~Figure 8).
%In this model, the stellar
%mass-to-light ratio $\Upsilon_{\star,B}$ is a factor of two lower than
%in the best fitting constant mass-to-light ratio model.
}

\noindent{{\bf Figure 10~}
Relative likelihood of all star+halo models that were
calculated in the 
parameter space of $\Upsilon_{\star,B}$ and $V_{200}$. For each model
the superposition of orbits was determined that best fits the NGC 2434
data. The area of each point is proportional to the logarithm of its
relative likelihood, which is determined by $\Delta\chi^2$. Models
with no dark halo ($V_{200} = 0$) are ruled out. A wide range of
potentials with a dark halo can fit the data. There is a strong
anti-correlation between $\Upsilon_{\star,B}$ and $V_{200}$ (see also
Figure 11); the
kinematics strongly constrain the mass within a given radius, but this
mass can either be attributed to the visible or the dark matter.}

\noindent{{\bf Figure 11~}
Confidence limits in the $\Upsilon_{\star,B}$ and $V_{200}$
plane for NGC 2434. The solid ellipse encloses the 1$\sigma$ (68\%)
confidence region, the dotted contour encloses the 2$\sigma$ (95.4\%)
confidence region, for the joint distribution of both parameters.
The solid and dotted lines denote the 1$\sigma$ and 2$\sigma$ limits
on the parameters individually: $\Upsilon_{\star,B}=3.35\pm0.25$
and $V_{200} = 450 \pm 100 \kms$.}

\begin{figure}
\epsscale{0.90}
\vskip -0.3truein
\plotone{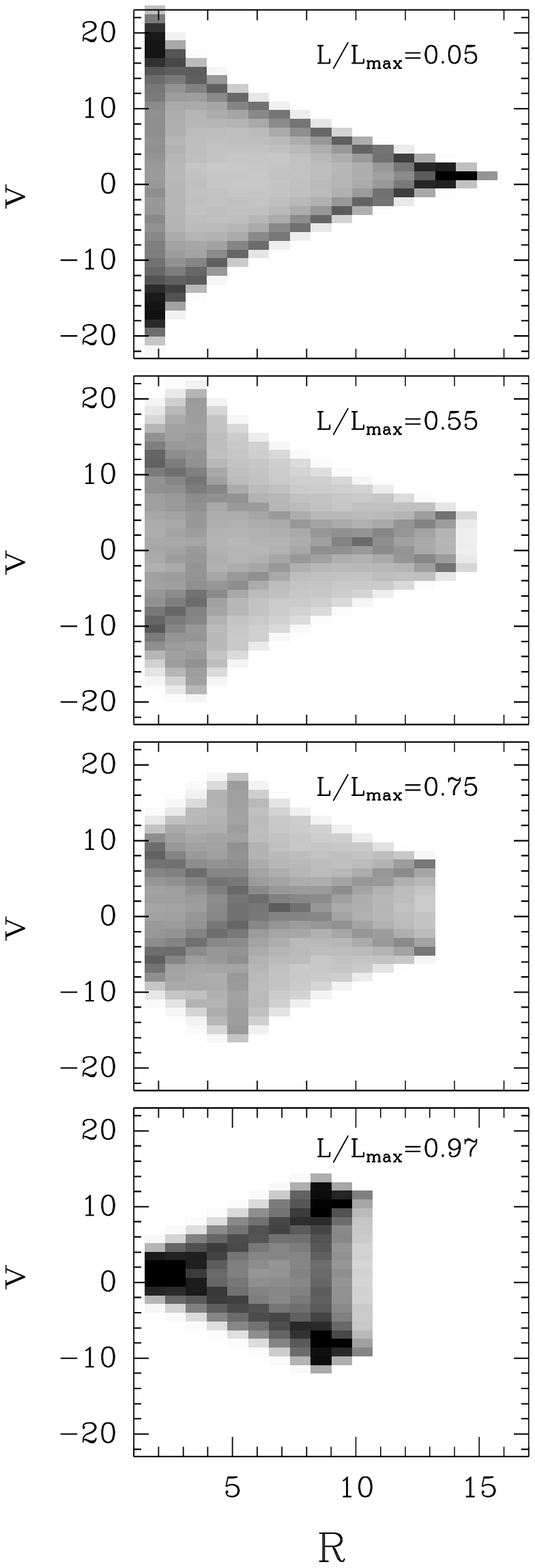}
\caption{ }
\end{figure}
\clearpage

\begin{figure}
\voffset-1.0truein
\epsscale{1.00}
\plotone{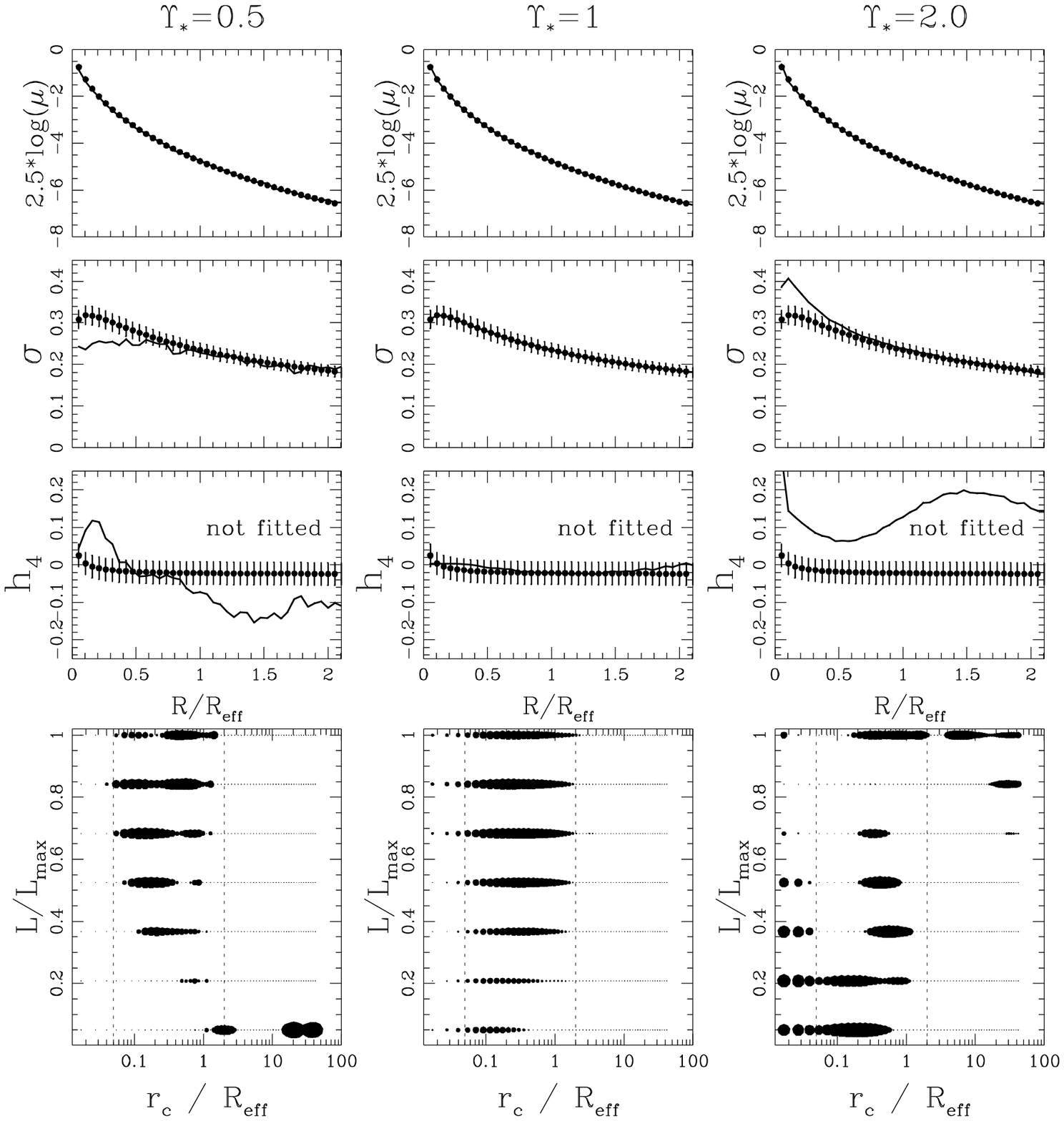}
\caption{ }
\end{figure}
\clearpage

\begin{figure}
\epsscale{1.00}
\plotone{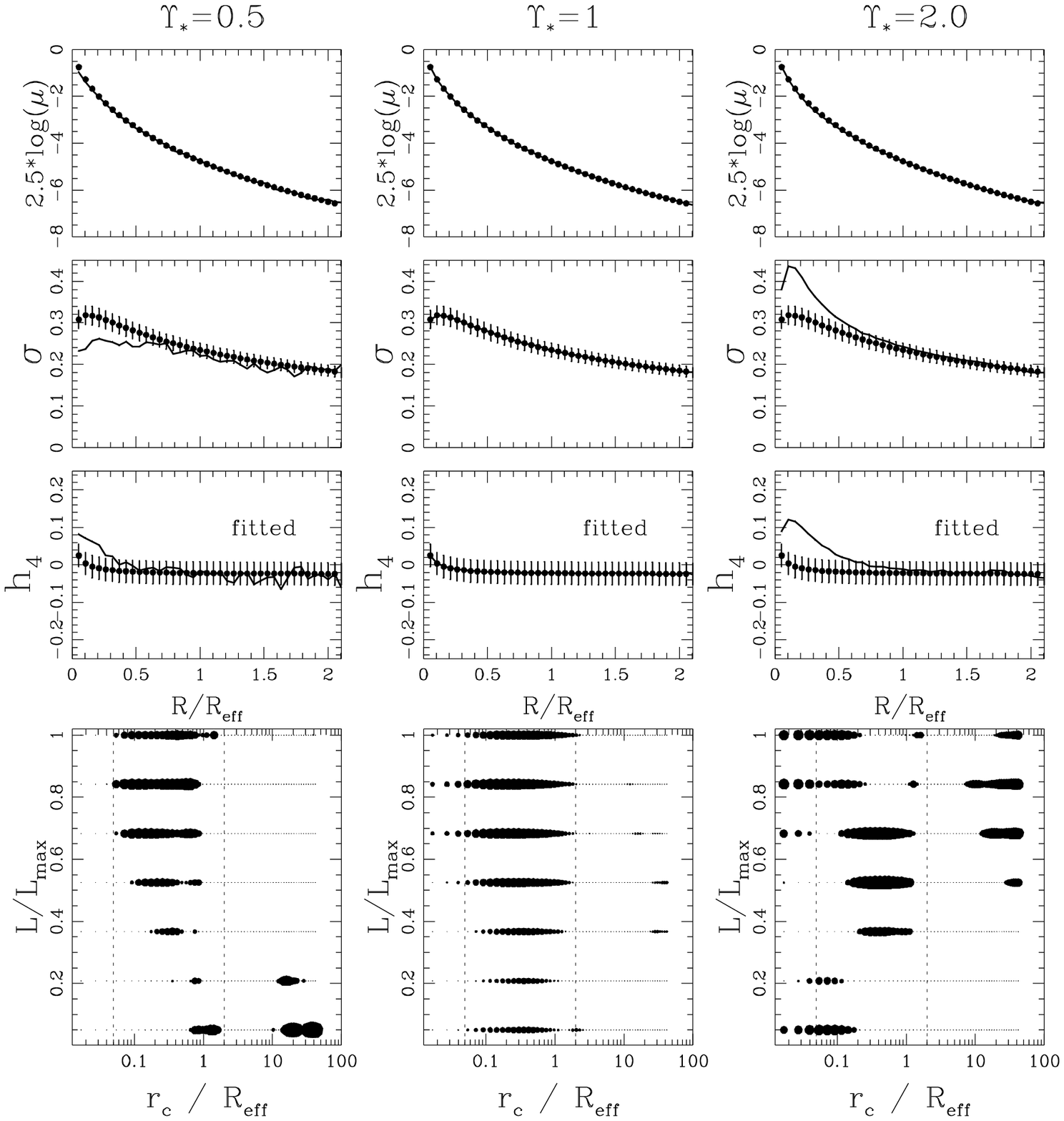}
\caption{ }
\end{figure}
\clearpage

\begin{figure}
\epsscale{1.00}
\plotone{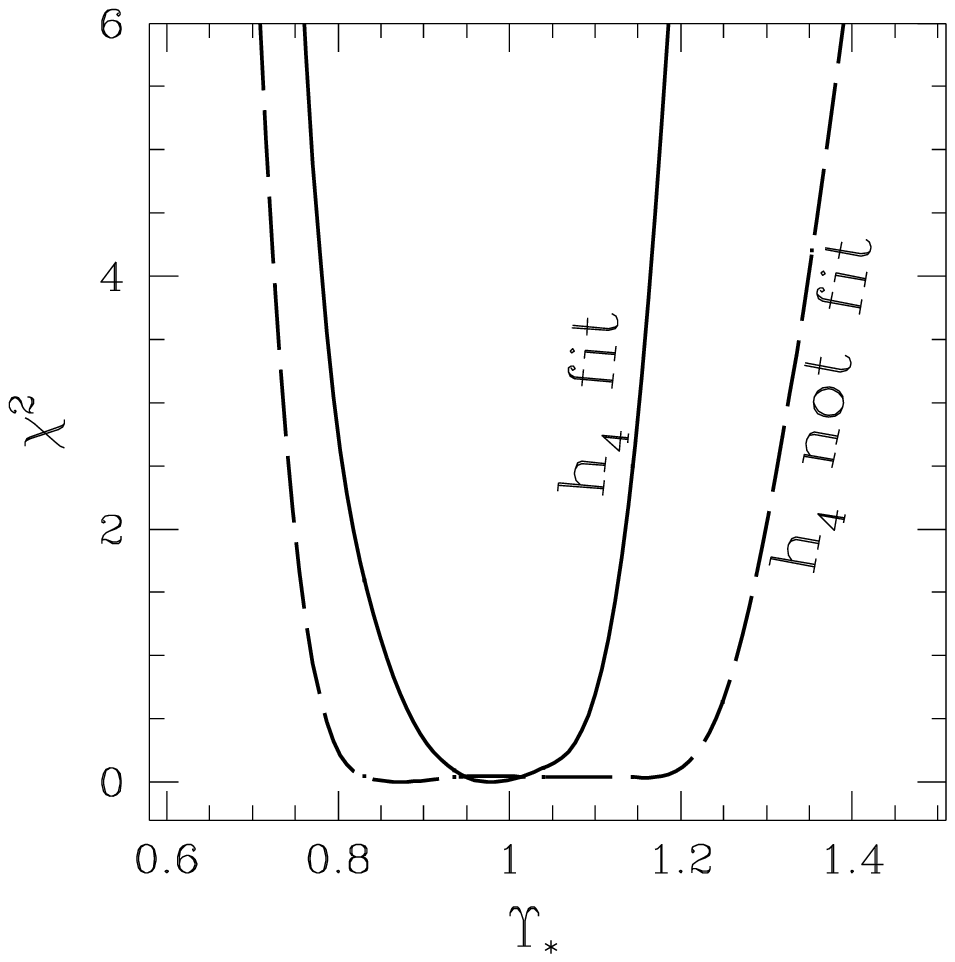}
\caption{ }
\end{figure}
\clearpage

\begin{figure}
\epsscale{1.00}
\plotone{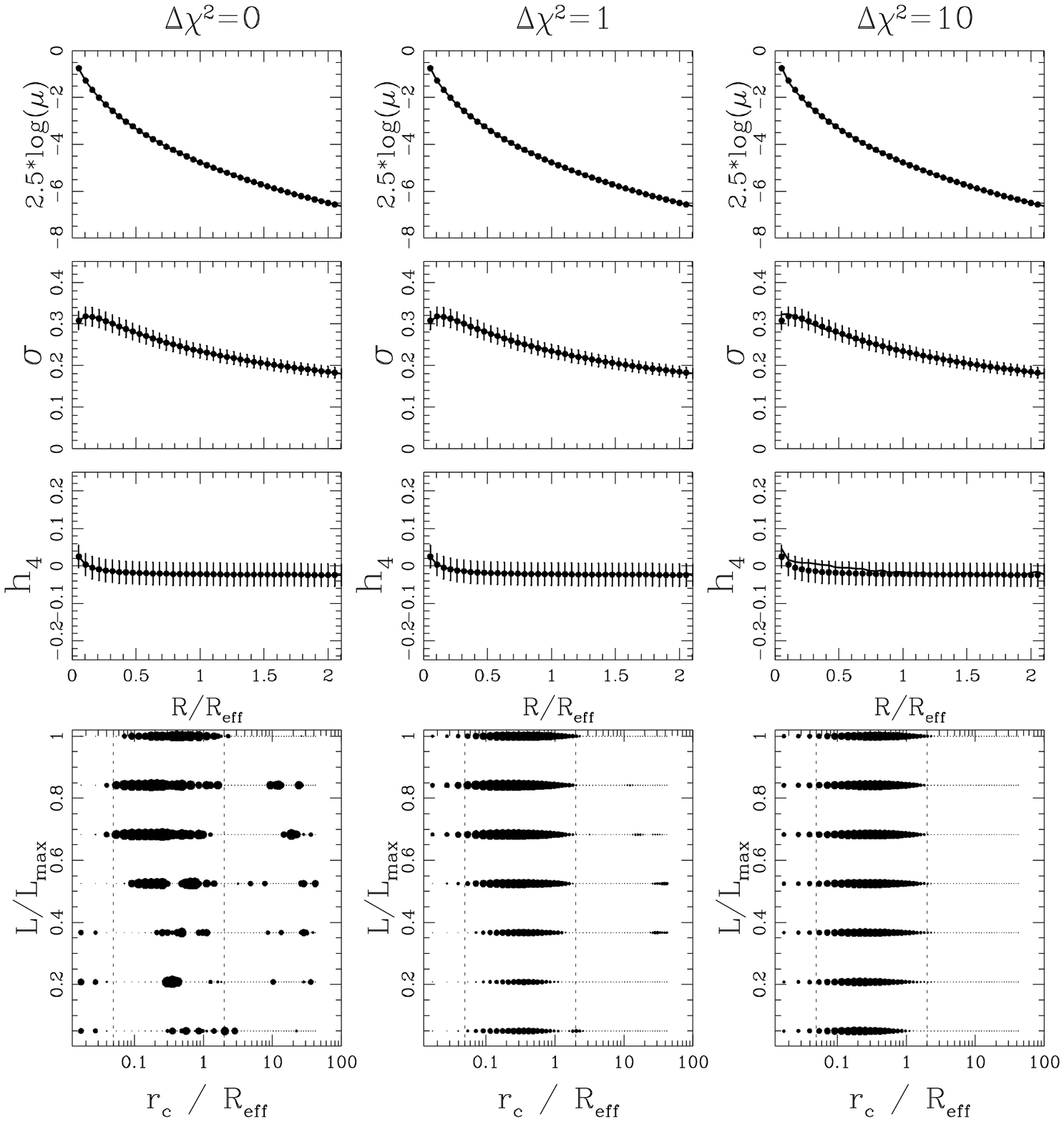}
\caption{ }
\end{figure}
\clearpage

\begin{figure}
\epsscale{1.00}
%\plotone{../N2434/Figure4.1.ps}
\plotone{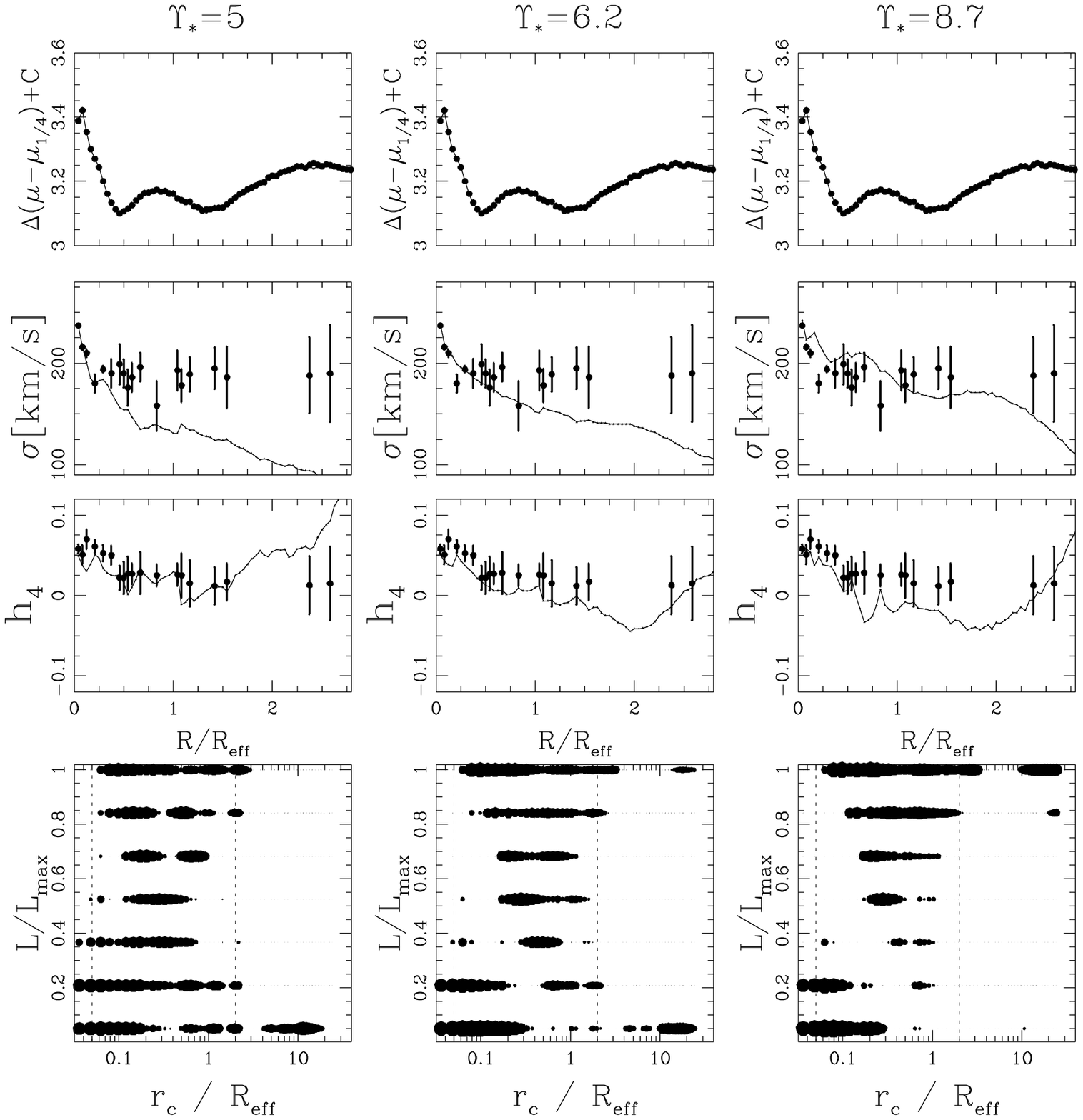}
\caption{ }
\end{figure}
\clearpage

\begin{figure}
\epsscale{1.00}
\vskip -1truein
\plotone{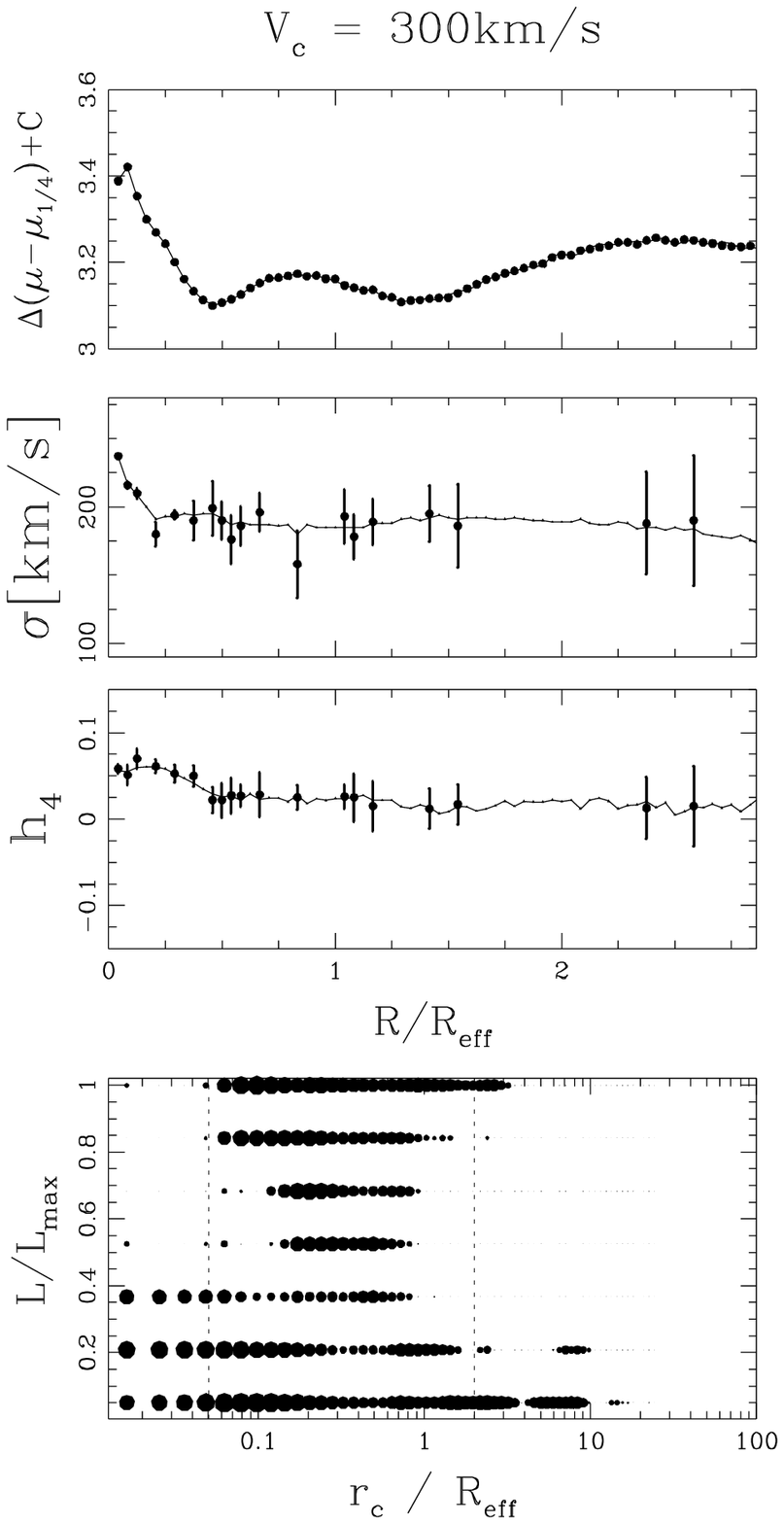}
\caption{ }
\end{figure}
\clearpage

\begin{figure}
\epsscale{1.00}
\plotone{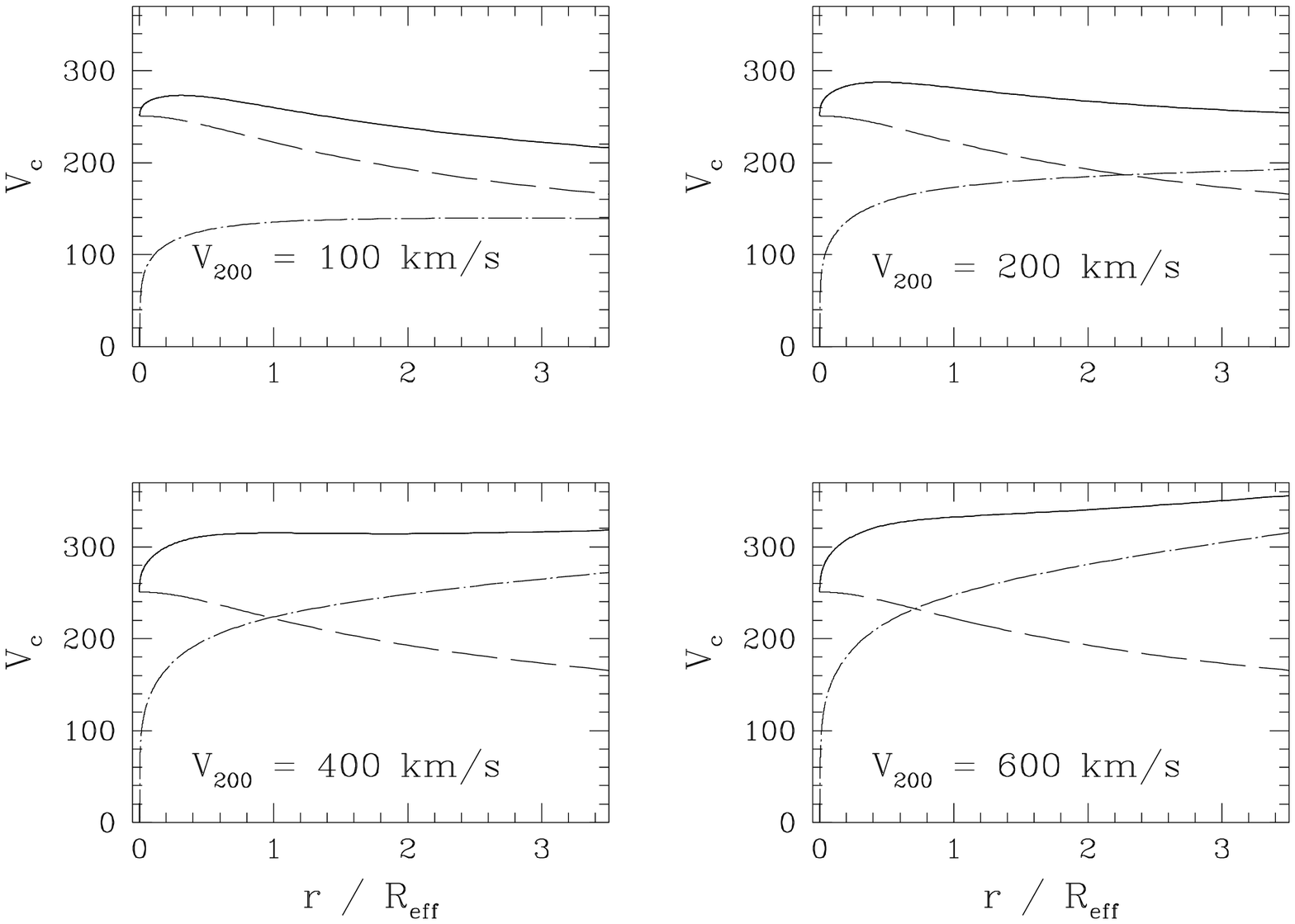}
\caption{ }
\end{figure}
\clearpage

\begin{figure}
\epsscale{1.00}
\plotone{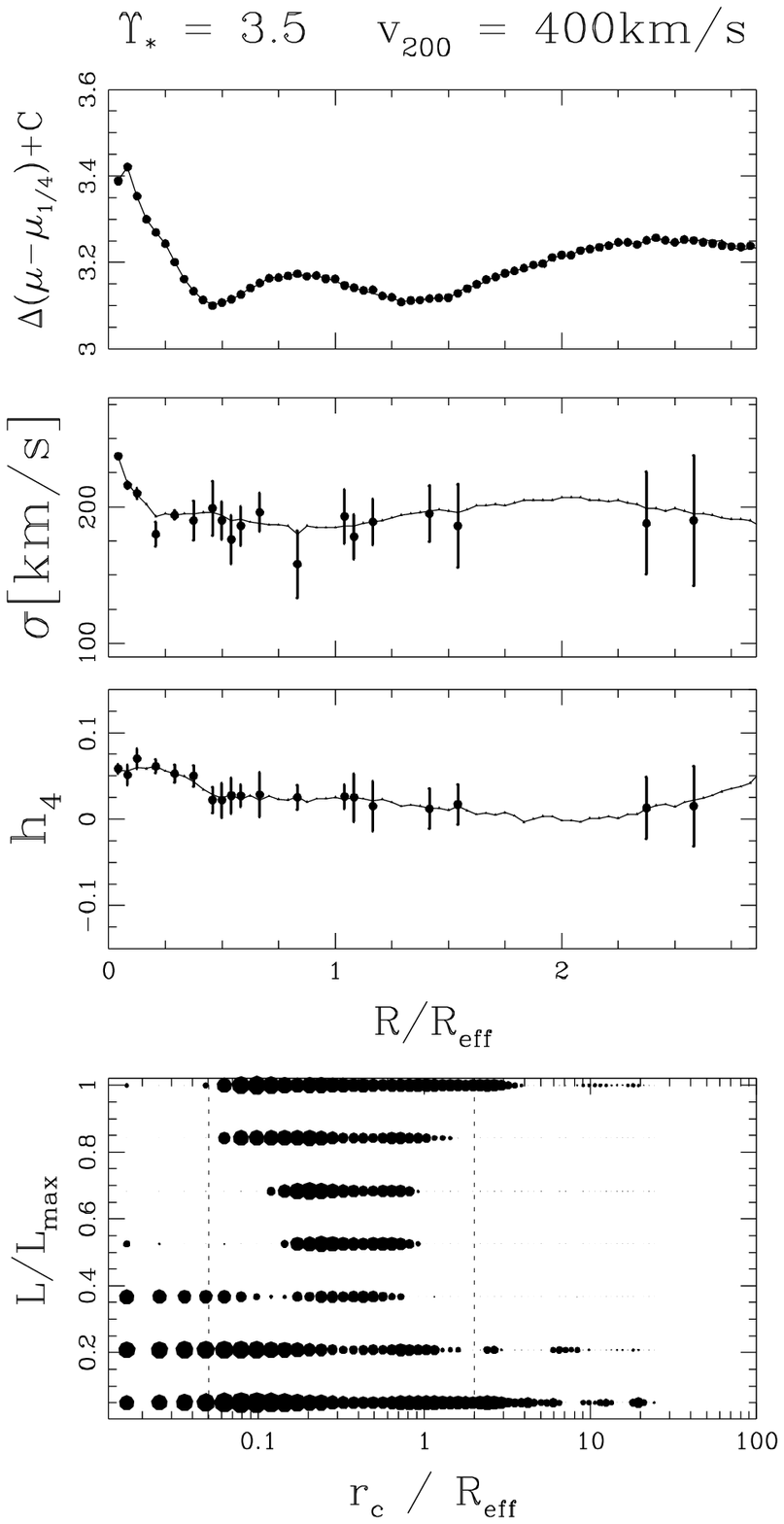}
\caption{ }
\end{figure}
\clearpage

\begin{figure}
\epsscale{1.00}
\plotone{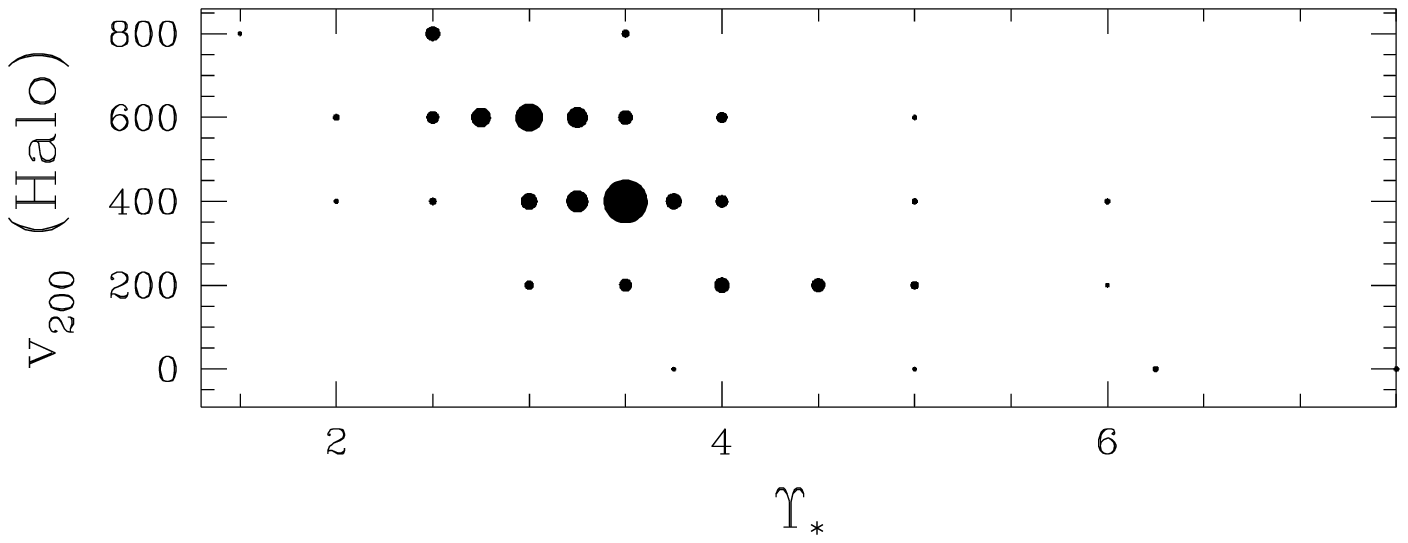}
\caption{ }
\end{figure}
\clearpage

\begin{figure}
\epsscale{1.0}
\plotone{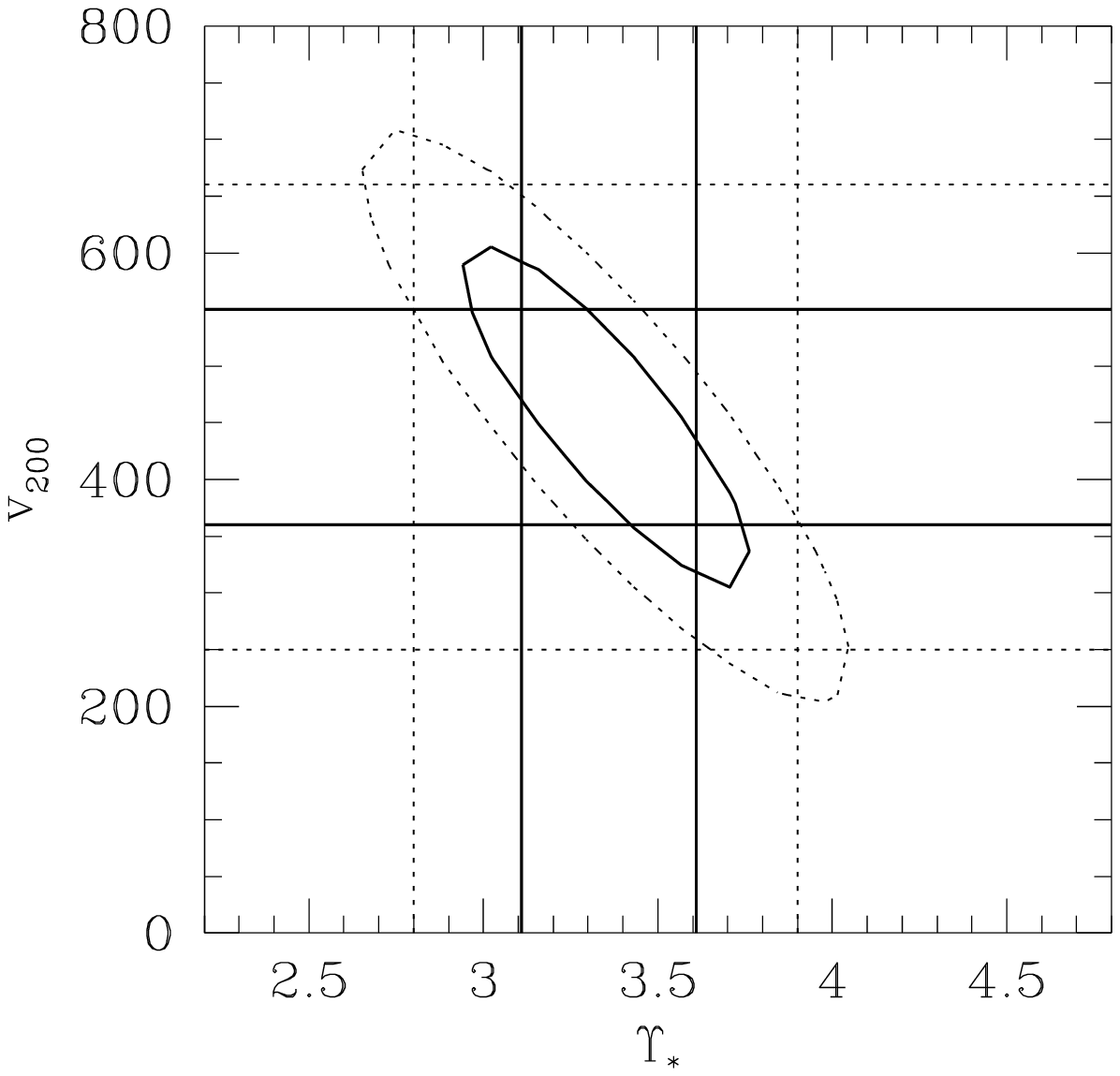}
\caption{ }
\end{figure}

\end{document}